\documentclass[12pt]{article}
\usepackage[pdftex]{graphics}

\pdfoutput=1

\newcommand{\preprint}[1]{\begin{table}[t]  
             \begin{flushright}               
             {#1}                             
             \end{flushright}                 
             \end{table}}                     
\renewcommand{\title}[1]{\vbox{\center\LARGE{#1}}\vspace{5mm}}
\renewcommand{\author}[1]{\vbox{\center#1}\vspace{5mm}}
\newcommand{\address}[1]{\vbox{\center\em#1}}

\usepackage{hyperref, amsmath, amssymb}
\usepackage{graphicx,color}
\usepackage{cite}
\usepackage{mciteplus}

\usepackage{subcaption}
\usepackage{epstopdf}
\usepackage{mathtools} 
\usepackage{float}
\numberwithin{equation}{section}

\newcommand\eq{\begin{equation}}
\newcommand\eqe{\end{equation}}
\newcommand\eqa{\begin{eqnarray}}
\newcommand\eqae{\end{eqnarray}}

\def\be#1\ee{\begin{align}#1\end{align}}

\begin{document}

\unitlength = .8mm

\begin{titlepage}

\begin{center}

\hfill \\
\hfill \\
\vskip 1cm

\title{Carving out the Space of Open-String S-matrix}

\author{Yu-tin Huang$^{1,2}$, Jin-Yu Liu$^1$, Laurentiu Rodina$^3$, Yihong Wang$^1$, }

\vspace{-0.35cm}
\address{$^1$Department of Physics, National Taiwan University, \\
No.1, Sec.4, Roosevelt Road, Taipei 10617, Taiwan\\
$^2$Physics Division, National Center for Theoretical Sciences,\\
National Tsing-Hua University,\\
$^3$
Institut de Physique Theorique, Universite Paris Saclay,
CEA, CNRS, F-91191 Gif-sur-Yvette, France
}

\preprint{NCTS-TH/2011}

\end{center}

\vspace{-0.30cm}

\begin{abstract}
\vspace{-0.10cm}
In this paper, we explore the open string amplitude's dual role as a space-time S-matrix and a 1D CFT correlation function.  We pursue this correspondence in two directions. First, beginning with a general disk integrand dressed with a Koba-Nielsen factor, we demonstrate that exchange symmetry for the factorization residue of the amplitude forces the integrand to be expandable on SL(2,R) conformal blocks. Furthermore, positivity constraints associated with unitarity imply the SL(2,R) blocks must come in linear combinations for which the Virasoro block emerges at the ``kink" in the space of solutions. In other words, Virasoro symmetry arises at the boundary of consistent factorization. Next, we consider the low energy EFT description, where unitarity manifests as the EFThedron in which the couplings must live. The existence of a worldsheet description implies, through the Koba-Nielsen factor, monodromy relations which impose algebraic identities amongst the EFT couplings.  We demonstrate at finite derivative order that the intersection of  the ``monodromy plane" and the EFThedron carves out a tiny island for the couplings, which continues to shrink as the derivative order is increased. At the eighth derivative order, on a three-dimensional monodromy plane, the intersection fixes the width of this island to around 1.5$\%$ (of $\zeta(3)$) and 0.2$\%$ (of $\zeta(5)$)  with respect to the super string answer. This leads us to conjecture that the four-point open superstring amplitude can be completely determined by the geometry of the intersection of the monodromy plane and the EFThedron.

\end{abstract}

\vfill

\end{titlepage}

\eject

\tableofcontents

\section{Introduction}
String theory amplitudes can be viewed as living in the intersection of two sets of consistency conditions: on the one hand they are subject to the usual analyticity constraints of the space-time S-matrix, on the other, as two-dimensional CFT correlators they must have a consistent operator product expansion (OPE). At four-points both sets of consistency systems are amenable to a bootstrap analysis, where a spectacular modern revival has been seen for the CFT bootstrap ~\cite{Rattazzi:2008pe}  (for a review see \cite{Simmons-Duffin:2016gjk, Poland:2016chs, Poland:2018epd}), and more recently for scattering amplitudes \cite{Paulos:2016fap, Paulos:2016but, Paulos:2017fhb, Homrich:2019cbt, Chowdhury:2019kaq, Bose:2020shm}. 

The interplay between space-time factorization and the structure of the worldsheet integrand has been one of the major undertones in the exploration of new representations of open string amplitudes. Indeed the fact that a string theory amplitude can be written as a product of a field-theory amplitude and its $\alpha'$ corrections can be viewed as manifesting its role as a UV completion of solutions to consistent massless factorization, as demonstrated for the superstring~\cite{Mafra:2011nv,Carrasco:2016ldy,Mafra:2016mcc} as well as bosonic and heteorotic strings~\cite{Huang:2016tag, Azevedo:2018dgo}. Even more recently, consistent factorization that extrapolates between the $\alpha'\rightarrow 0$ limit and finite $\alpha'$ naturally led to the realization of the integrand as stringy-canonical forms~\cite{Arkani-Hamed:2019mrd}. As these features stem from considering massless poles, it is natural to ask what is the image of consistent \textit{massive} factorization on the worldsheet integrand. 

On the other hand, UV unitarity leads to non-trivial bounds on the low energy effective field theory (EFT) couplings. This was famously explored for the positivity of the leading four-derivative coupling stemming from the optical theorem in~\cite{Adams:2006sv}. Extensions to higher order derivatives, and away from the forward limit, were explored in subsequent works \cite{Bellazzini:2015cra, deRham:2017avq, deRham:2017xox, Chen:2019qvr}, with a complete geometric definition identified as the EFThedron~\cite{EFThedron}. These are general unitarity constraints that do not require the presence of a worldsheet, and were studied recently for string amplitudes in \cite{Green:2019tpt}. In this work, we would like to ask if there is any feature of the worldsheet that has a distinctive projection in the EFThedron. 

In this paper, we wish to study the projection in both directions. The arena is the following ansatz for an open string amplitude
\eq\label{IntroA}
A(s,t)\sim\int_0^1 dz\; z^{\alpha' 2k_3\cdot k_4}(1-z)^{\alpha' 2k_2\cdot k_3}f(z)\, ,
\eqe
where $f(z)$ is some function that is analytic near $z=0$, and $k_i$ are $d$-dimensional momenta. We use $s={-}(k_1{+}k_2)^2$, $t={-}(k_1{+} k_4)^2$, and $u={-}(k_1{+} k_3)^2$, so that $A(s,t)\equiv A(1234)$ is an ordered amplitude with only $s$ and $t$ channel poles. This can be viewed as the amplitude for the vacuum state of the compactified string on a product geometry $R^{1,d{-}1}\otimes M$, where $M$ is compact. 

\noindent{\textbf{Space-time constraints on the worldsheet}}

\begin{figure}
  \begin{center}
\includegraphics[scale=0.5]{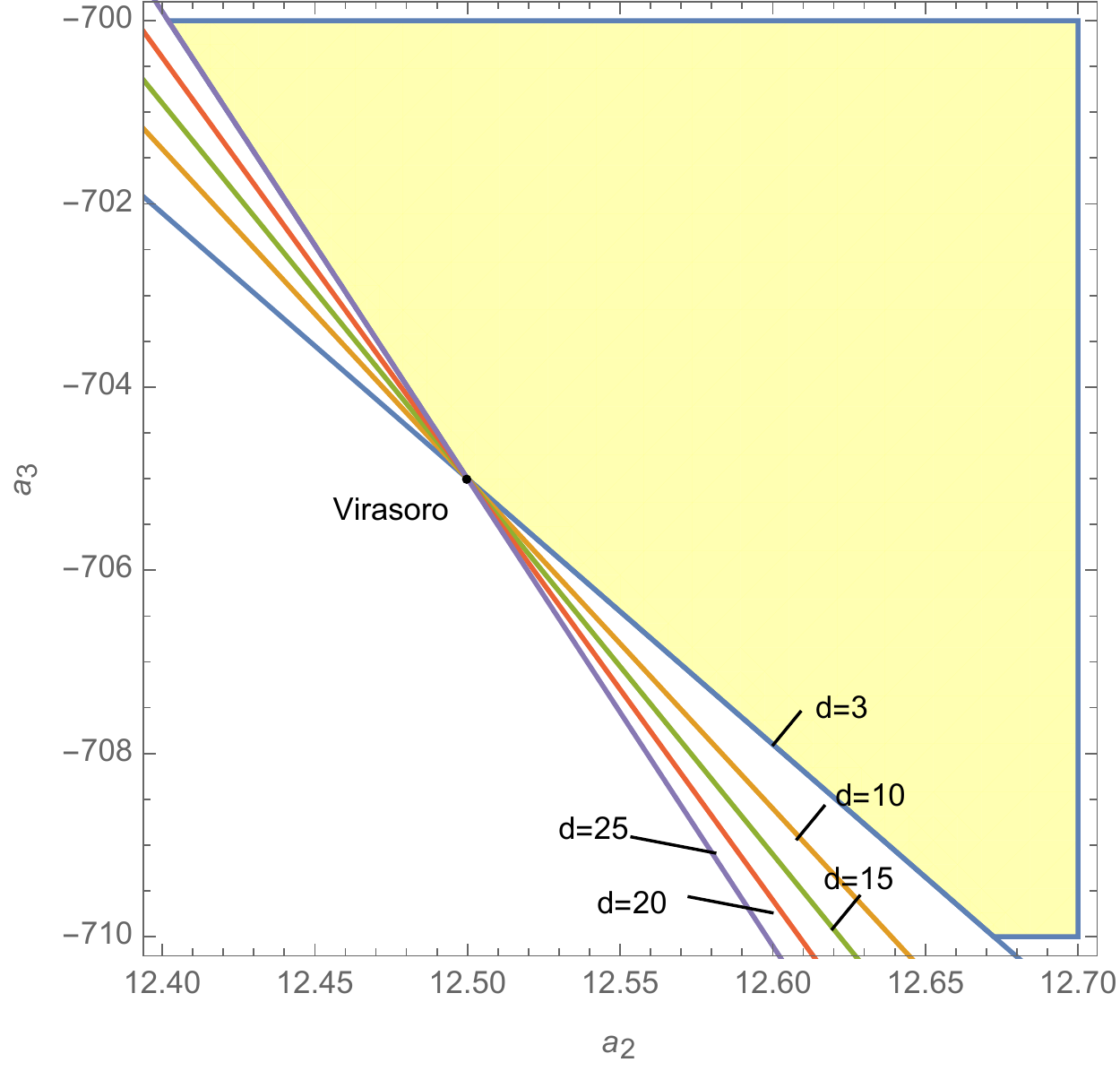}
        \caption{The plot for the allowed solution for an ansatz for $\chi_4=\frac{(i+a_4 i^2) (j+a_4 j^2)}{a_1 d^2 +a_2 d+a_3}$ under (ii) and (iii). For illustrative purposes we have set  $a_4=5, a_1=9880$, the Virasoro values. The lines denote the boundaries carved out by (iii) as a function of the space-time dimension. The point in the figure is the Virasoro value for $a_2$ and $a_3$.}
          \label{2DPlot0}
     \end{center}
\end{figure}

We first consider constraints on the function $f(z)$ imposed by the fact that $A(s,t)$, as a space-time S-matrix, must factorize in a way that is consistent with unitarity and Lorentz symmetry. The latter implies that the residue must be expandable on the Gegenbauer polynomials\footnote{Recall that the Gegenbauer polynomials are orthogonal polynomials that form irreducible representations of SO($d{-}1$). They are given by the following generating function
\eq\label{GegGen}
\frac{1}{(1-2r\cos\theta+r^2)^{\frac{d-3}{2}}}=\sum_{\ell}\; r^\ell G_\ell^d(\cos\theta)
\eqe
} 
\eq
\textrm{Res}[A(s,t)]\bigg|_{s\rightarrow m^2}=\sum_\ell \mathcal{C}_\ell\frac{G_\ell^d(\cos\theta)}{s-m^2}\,,
\eqe
while the former implies linear and quadratic bounds on $\mathcal{C}_\ell$. In particular, labeling the mass of the external legs as $i_1, i_2, i_3$ and $i_4$, unitarity  implies 
\begin{itemize}
  \item (i) $\mathcal{C}_{\ell} $ is symmetric  under $i_1 \leftrightarrow i_2 $, $i_3 \leftrightarrow i_4$ exchange.
  \item (ii) $\mathcal{C}_{\ell} $ must be positive when  $i_1=i_4$ and $i_2=i_3$. 
  \item (iii) For distinct masses, the  $\mathcal{C}_{\ell}$'s satisfy the Schwarz inequality:
  \eq
 \mathcal{C}_{\ell}(i_1, i_2,i_2,i_1)\mathcal{C}_{\ell}(i_3, i_4,i_4,i_3)- \left(\mathcal{C}_{\ell}(i_1, i_2,i_3,i_4)\right)^2 \geq 0\,.
  \eqe 
\end{itemize}
Since $f(z)$ is analytic near the origin, we can easily extract the $s$-channel factorization residues and consider the consequences of the above constraints on the power series coefficients of $f(z)$. We find that (i) alone is sufficient to show that $f(z)$ must be given by linear combinations of SL(2,R) global conformal blocks, expressed as hypergeometric functions of the type $_2F_1(\Delta,\Delta, 2\Delta, z)$. Conditions (ii) and (iii) impose further constraints. In particular, we will show that for these constraints to hold for arbitrary positive external masses, the global blocks whose conformal dimensions differ by integers must be grouped into a new function of the form:
\eq
z^{\Delta} \,_2F_1(\Delta,\Delta, 2\Delta, z) +\sum_{q=1}^\infty \chi_q z^{\Delta{+}q} \,_2F_1(\Delta{+}q,\Delta{+}q, 2\Delta{+}2q, z)\,,
\eqe
with their relative coefficients $\chi_q$ bounded by (ii) and (iii). Using this criteria we find that $\chi_i$ with $i=1,2,3$ can be completely determined as the solution that saturates the bounds in (ii) and (iii). For $\chi_4$ where things become more subtle, we can plot the allowed region and demonstrate that the Virasoro block lives at the kink of the region, as shown in Figure (\ref{2DPlot0}). Thus we see that worldsheet Virasoro symmetry emerges at the boundary of consistent space-time factorization.\\
\\

\noindent{\textbf{Worldsheet image on space-time S-matrix}}

We then consider the projection of constraints in the opposite direction. Starting from (\ref{IntroA}),  the Koba-Nielsen factor implies the following monodromy relation amongst amplitudes of different orderings~\cite{Stieberger:2009hq,BjerrumBohr:2009rd}: \begin{equation}
A\left(s,u\right)+e^{i\pi  s}A\left(s,t\right)+e^{-i\pi u }A\left(t,u\right)=0 \,.
\label{monodromy}
\end{equation}
Taking the low energy limit and expanding the amplitude in Mandelstam invariants, 
\eq
A(s,t)|_{s,t\ll 1}=({\rm massless\,poles})+\sum_{k,q\geq 0} g_{k,q}s^{k{-}q}t^q\,,
\eqe
the monodromy relation imposes algebraic identities between the EFT couplings $g_{k,q}$, which importantly can also be amongst couplings of different mass dimensions. For instance, it fixes $g_{0,0}=\frac{\pi^2}{6}$, or $g_{3,1}=2g_{3,0}-\frac{\pi^2}{6}g_{1,0}$. The remaining free parameters, like $g_{1,0}$ or $g_{3,0}$, define the ``monodromy plane" in the space of EFT couplings. 

\begin{figure}
  \begin{center}
\includegraphics[scale=0.5]{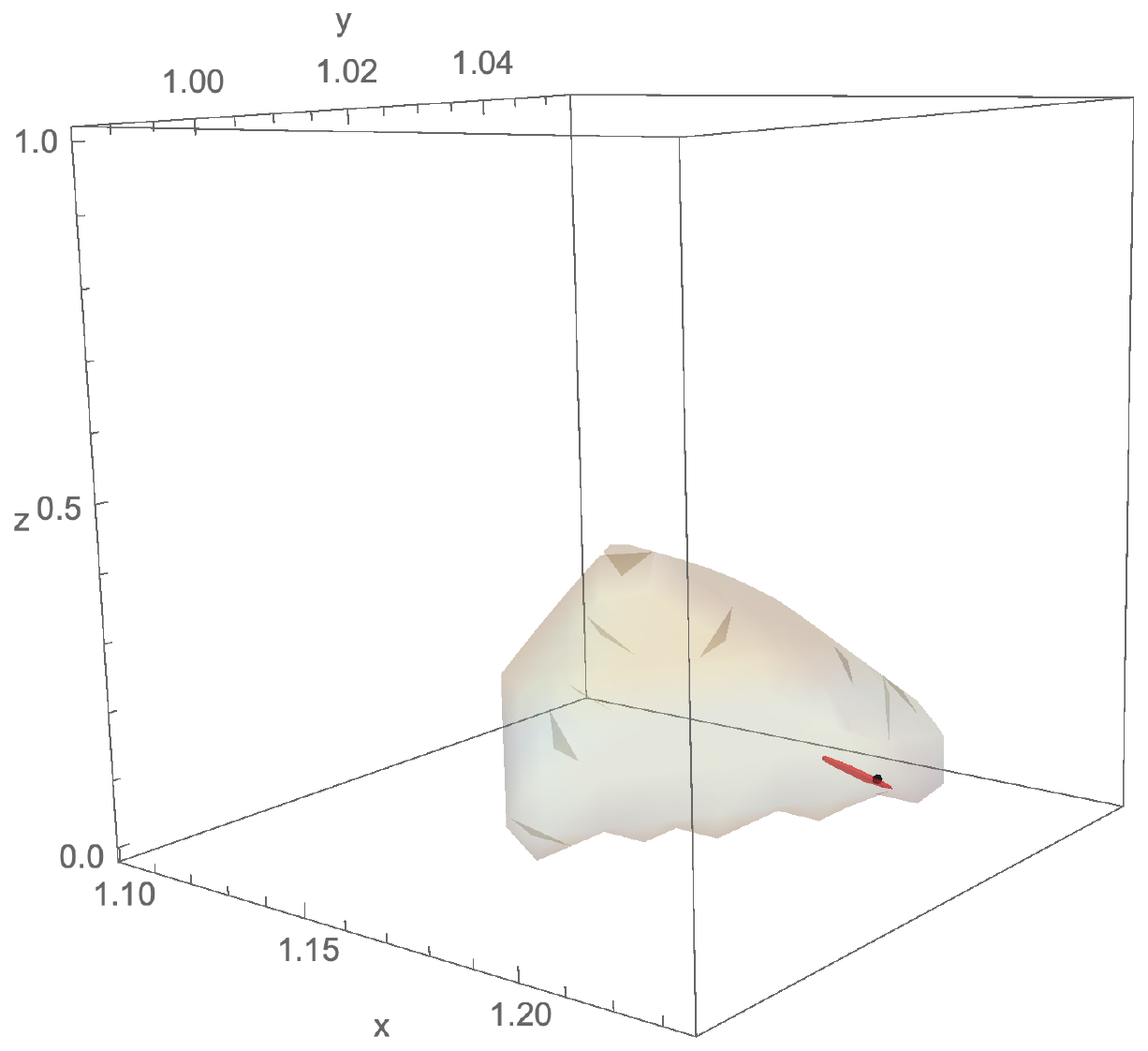}
        \caption{The gray region represents the three-dimensional intersection of the monodromy plane and EFThedron at eighth derivative order. The red region represents the region that can be projected from the four-dimensional geometry that would appear at ninth derivative order.    
        }
      
          \label{IntroPlot2}
     \end{center}
\end{figure}

On the other hand, UV unitarity, Lorentz invariance and locality also constrain the space of allowed couplings to be inside the EFThedron~\cite{EFThedron}. Thus the image of the worldsheet inside the EFThedron is given by its intersection with the monodromy plane. Remarkably, we find that the intersection gives just a tiny allowed region for the independent EFT couplings. For example, up to $k=4$ (eight derivative order), the monodromy plane is three-dimensional and parameterized by $g_{1,0}$, $g_{3,0}$ and $g_{4,1}$. Applying EFThedron constraints on this space, we obtain a finite intersection region displayed in gray in Figure \ref{IntroPlot2}. The region is drastically reduced by requiring it to be uplifted into the $k=6$ geometry, for the cases when  the four-dimensional monodromy plane intersects with the EFThedron. The reduced region is displayed as red, and fixes the coefficients with the following precision:
\eqa
&&\nonumber\frac{g_{1,0}^{max}-g_{1,0}^{min}}{g_{1,0}^{string}}=\frac{1.20667-1.18890}{\zeta(3)}\approx1.5\%\,, \\
&&\frac{g_{3,0}^{max}-g_{3,0}^{min}}{g_{3,0}^{string}}=\frac{1.03808-1.03594}{\zeta(5)}\approx0.2\%\,,\nonumber\\
 &&\frac{g_{4,1}^{max}-g_{4,1}^{min}}{g_{4,1}^{string}}=\frac{0.05699-0.03560}{(\pi^6-630\zeta(3)^2)/1260)}\approx52.8\%\,.
\eqae
Applying all $k=7$ and one $k=8$ constraints, using the \textbf{FindInstance} function in \emph{Mathematica} we were able to further shrink to region in Figure  \ref{Remain}. In fact, by setting $(g_{1,0},g_{3,0},g_{4,1})$ to string values $(\zeta(3),\zeta(5),(\pi^6-630\zeta(3)^2)/1260)$, we can search for solutions to the $k=7$ EFThedron constraints using \textbf{FindInstance}, finding:
\eq
(g_{5,0},g_{6,1},g_{7,0},g_{7,2})=(1.00834,0.00862,1.00202,0.00035)\,,
\eqe 
which matches to string values up to four digits:
\eqa
\nonumber&&\left(\zeta(7),\frac{\pi^8}{7560}{-}\zeta(3)\zeta(5),\zeta(9),\frac{8\pi^6\psi^{(2)}(1){+}9\pi^4\psi^{(4)}(1){+}6\pi^2\psi^{(6)}(1){-}180\psi^{(2)}(1)^3{-}2\psi^{(8)}(1)}{8640}\right)\\
&&=(1.00835,0.00865,1.00201,0.00032)\,.
\eqae
This analysis leads us to the conjecture that:

\vspace{5pt}
 \textit{The geometry of intersection between the monodromy plane and the EFThedron yields the four-point massless amplitude of Type-I superstring}.
\vspace{5pt}

Finally, note that (\ref{monodromy})  is the monodromy relation for amplitudes with identical external states, therefore can only be satisfied by (\ref{IntroA}) when $f\left(z\right)$ is a symmetric function modulo $SL(2)$. A non-symmetrical $f\left(z\right)$ may lead to variations of (\ref{monodromy}), characterizing the permutational behaviour of the external states.  We repeat the monodromy and unitarity study on one of such variation: the bicolour monodromy relation  $\frac{u}{s}A\left(s,u\right)+e^{i \pi s}A\left(s,t\right)+e^{-i\pi u}\frac{t}{s}A\left(u,t\right)=0$, and achieve results similar to the single color amplitude. 

This paper is organized as follows. In Section 2 we discuss the emergence of global conformal and Virasoro blocks from unitarity. Section 3 is a review of the EFThedron, the positive geometry in which EFT parameters must exist in order to satisfy unitarity. In Section 4 we solve the monodromy condition perturbatively, and extract the relations imposed between different physical parameters, the monodromy plane. In Section 5 we apply EFThedron considerations on the remaining free parameters of the monodromy plane, and demonstrate how considering higher and higher order cyclic polytopes and Hankel matrices drastically reduces the allowed physical space, apparently converging to the open string amplitude. We end with conclusions and future directions in Section 6.

\section{Consistent factorization and the emergence of Virasoro symmetry} 
In this section we ask the following question: suppose we have a four-point scalar amplitude that takes the form 
\eqa\label{MainAnsatz1}
\label{dap}A(s,t)&=&\int_0^1 dz\; z^{\alpha' 2k_3\cdot k_4}(1-z)^{\alpha' 2k_2\cdot k_3}f(z)\,,
\eqae
where $\alpha'$ is a normalization scale and we assume $f(z)$ is analytic in the positive region near $z=0$. What constraints must $f(z)$ respect to ensure the consistency of the space-time $S$-matrix $A(s,t)$? The form of (\ref{MainAnsatz1}) can be motivated from several fronts. Firstly, the kinematic dependence is completely contained in the Koba-Nielsen factor, which leads to exponential softness at $s,{-}t\gg 1$ \`a la Gross and Mende~\cite{Gross:1987kza} (see~\cite{Gross:1989ge} for open strings), as well as a linear trajectory for $s, t\gg 1$, which was shown to be universal in \cite{Caron-Huot:2016icg}. Secondly, one can consider this as an ansatz for the scattering of the vacuum state in $d$-dimensions, for string theory compactified on $R^{1,d{-}1}\otimes M_{d_c{-}d}$, where $d_c$ is the critical dimension.  

From the CFT perspective, we expect $f(z)$ to be a four-point correlation function with an OPE expansion, schematically given by 
\begin{align}
\langle  O_1O_2 O_3 O_4\rangle \propto \sum_p C_{12p}C_{p34} \mathcal{F}(h_i,h_p,z)\,,
\end{align}
where the $C$'s are OPE coefficients, and $\mathcal{F}$'s are the conformal blocks, representing the exchange of either SL(2,R) or Virasoro primaries. These blocks have distinct series expansions near  $z\rightarrow0$. Thus we would like to see how consistency of the space-time S-matrix requires the $f(z)$ to have an expansion on the blocks $\mathcal{F}$. 

Since the kinematic dependence is all in the Koba-Nielsen factor, the amplitude can only develop singularities when this factor diverges, i.e. at the boundary of the integration region $z=0,1$. In other words, we will only have $s$ and $t$-channel singularities, implying that the amplitude has a prescribed ordering. This motivates us to re-express (\ref{MainAnsatz1}) as 
\eqa\label{MainAnsatz2}
A(s,t)&=&\int_0^1 dz\; z^{-s +i_3+i_4-2}(1-z)^{ t+i_2+i_3-2}f(z)\,,
\eqae
where we denote the mass of each leg as $m^2_a=i_a-1$, where the $-1$ at this point is just convention. Note that we have set $\alpha'=1$. Since $f(z)$ is analytic in the positive region near the origin $z=0$, we can extract the residue of the $s$-channel singularity, by writing $f(z)$ as a power series with generic real exponents (Hahn series) in the neighborhood of $z=0$, which without loss of generality we write as $f (z)|_{z\rightarrow 0}=\sum c_p z^{p-i_3-i_4}$. This can be viewed as a sum of ``dilatation blocks". For each individual block, the integral 
\eqa
\int_0^1 dz\; z^{-s-2}(1-z)^{ t+i_2+i_3-2} z^{p}\,,
\eqae
will have $s$-channel poles at $p-1, p, p+2,\cdots$. Since we will be imposing consistency conditions on the factorization poles, it is natural to collect the dilatation blocks that differ by integers into a subset. Thus we consider 
\eq
f_p (z)=z^{p-i_3-i_4}D_{\{i_a\}},\quad D_{\{i_a\}}=1+v_1 z+ v_2 z^2+\cdots =1+\sum_{i=1}^\infty v_iz^i\,.
\eqe
Note that the relative coefficients $v_i$ are understood to be functions of $\{i_a,p\}$.  The $s$-channel residue is then given by evaluating the integral in (\ref{MainAnsatz2}) as a contour integral around $z=0$ for fixed $s$. We see that the $s$-channel singularity appears at $s=n+p-1$, where $n$ is a non-negative integer, and the residue is given by
\eq\label{ResDef}
\left.\frac{d^n}{dz^n}(1-z)^{t+i_2+i_3-2}D_{\{i_a\}}\right|_{z=0}\equiv \textrm{Res}_n(t)\,.
\eqe
Since $A(s,t)$ is a $d$-dimensional scalar scattering amplitude, we expand the residue function $\textrm{Res}_n(t)$ in the Gegenbauer polynomial basis. This can be done by first converting $t$ to the center of mass scattering angle via:
\eqa
\cos\theta&=&\frac{(s+m_1^2-m_2^2)(s+m_4^2-m_3^2)-2s(m_1^2+m_4^2-t)}{\sqrt{(s-m_1^2-m_2^2)^2-4m_1^2m_2^2}\sqrt{(s-m_3^2-m_4^2)^2-4m_3^2m_4^2}}\,,
\eqae  
where $s$ is to be evaluated at $s=n+p-1$. Using this we can expand the residue on the $d$-dimensional Gegenbauer polynomials $G^{d}_\ell (\cos\theta)$:  
\eqa
\textrm{Res}_n(\cos\theta)&=&\sum_{\ell} \mathcal{C}^{(n)}_{\ell} K^{\ell} G^{d}_\ell (\cos\theta)\,,
\eqae
where $K$ is a kinematic factor given as:
\eq
K=\frac{\sqrt{(n{+}p{-}m_1^2{-}m_2^2{-}1)^2{-}4m_1^2m_2^2}\sqrt{(n{+}p{-}m_3^2{-}m_4^2{-}1)^2{-}4m_3^2m_4^2}}{n{+}p{-}1}\,.
\eqe
For example, expanding (\ref{ResDef})  onto the above basis up to level $2$, the $\mathcal{C}^{(n)}_{\ell}$ are given by:\footnote{At level $n=0$ we only have $\mathcal{C}^{(0)}_{0}$, which is simply a constant proportional to $f(0)$. }
\eqa\label{Cexamp}
\mathcal{C}_0^{(1)}&=&{-}\frac{({-}i_1 {+} i_2 {+} p) (i_3 {-} i_4 {+} p)}{2 p} {+} v_1\, ,\\
\quad \mathcal{C}_1^{(1)}&=&\frac{1}{2 ({-}3 {+} d)}\,,\\
\nonumber\mathcal{C}_0^{(2)}&=&\frac{d(p{+}1)^2}{8(d{-}1)}{+}\frac{d(i_1{-}i_2)^2(i_3{-}i_4)^2}{8(d{-}1)(p{+}1)^2}{-}\frac{(p{+}1)(d{-}5{+}d(i_1{-}i_2{-}i_3{+}i_4){+}2(i_2{+}i_3))}{4(d{-}1)}\, ,\\
\nonumber&&{+}\frac{d}{8(d{-}1)}(i_1^2{-}2 i_2 i_1{-}4 i_3 i_1{+}4 i_4 i_1{+}2 i_1{+}i_2^2{+}i_3^2{+}i_4^2{-}2 i_2{+}4 i_2 i_3{-}2 i_3{-}4 i_2 i_4{-}2 i_3 i_4{+}2 i_4)\\
\nonumber&&{+}\frac{1}{4(d{-}1)}(4 i_3 i_1{-}5 i_1{-}3 i_2{-}3 i_3{+}4 i_2 i_4{-}5 i_4{+}8)\\
\nonumber&&{+}\frac{1}{4 (d{-}1) (p{+}1)}(i_1^2 ((d{-}2) i_3{-}d i_4{+}2){-}i_1 (2 i_2 ((d{-}2) i_3{-}d i_4{+}2){+}(i_3{-}i_4) (d i_3{-}d i_4{-}d{+}1))\\
\nonumber&&{+}i_2 (i_3{-}i_4) ((d{-}2) i_3{-}(d{-}2) i_4{-}d{+}1){+}i_2^2 ((d{-}2) i_3{-}d i_4{+}2){+}2 (i_3{-}i_4)^2)\\
&&{+}\frac{(i_1{-}i_2{-}p{-}1) (i_3{-}i_4{+}p{+}1)}{2 (p{+}1)}v_1{+}v_2\, ,\\
\mathcal{C}_1^{(2)}&=&{-}\frac{(i_2{-}i_1 )(i_3{-}i_4{+}p{+}1) {+}(p{+}1) (i_3{-}i_4{+}p)}{4 (d{-}3)(p{+}1)}{+}\frac{1}{2(d{-}3)}v_1\, ,\label{Cexamp2}\\
\mathcal{C}_2^{(2)}&=&\frac{1}{4 (d{-}3)(d{-}1)}\, .\label{Cexampf}
\eqae
Thus we see that the coefficients $\mathcal{C}^{(n)}_{\ell} $ will in general be given as functions of $\{i_a, d, p\}$, and the unknown Taylor coefficients $v_i$ of $D_{\{i_a\}}$. For fixed $n$, they are non-vanishing for $\ell\leq n$, and their dependence on $v_i$ is given as: 
\begin{center}
    \begin{tabular}{ | c | c | c | c | c |}
    \hline
    Level-n  & Spin-0 & Spin-1 & Spin-2 & Spin-3  \\ \hline
    0 & $(v_0)$ & 0 & 0 & 0  \\ \hline
    1 & $(v_0,v_1)$ & $(v_0)$ & 0 & 0  \\ \hline
    2 & $(v_0,v_1,v_2)$ & $(v_0,v_1)$ & $(v_0)$ & 0  \\ \hline
    3 & $(v_0,v_1,v_2,v_3)$ & $(v_0,v_1,v_2)$  & $(v_0,v_1)$ & $(v_0)$  \\
    \hline
    \end{tabular}
\end{center}
$v_0$ is just a normalization of $f_p (z)$, which we set to $1$.

Now since the residue function $\textrm{Res}_n(t)$ must have an interpretation as the factorization of the four-point amplitude into the product of three-point amplitudes, it should have an equivalent representation as
\eq\label{AmplitudeFac}
\textrm{Res}_n(t)=\sum_{\ell}\quad g^{i_1 i_2}_\ell  k^{\mu_1}_{12}\,k^{\mu_2}_{12}\,\cdots k^{\mu_\ell}_{12}\; \mathbf{P}_{\mu_1\mu_2\cdots \mu_\ell;\nu_1\nu_2\cdots \nu_\ell}\; k^{\nu_1}_{34}\,k^{\nu_2}_{34}\,\cdots k^{\nu_\ell}_{34} g^{i_3 i_4}_\ell \,,
\eqe
where $k^\mu_{12}=k_1^\mu-k_2^\mu$ and $\mathbf{P}_{\mu_1\mu_2\cdots \mu_\ell;\nu_1\nu_2\cdots \nu_\ell}$ is a degree $\ell$ polynomial in $\eta^{\mu\nu}$, that is symmetric and traceless in $\{\mu\}$ and $\{\nu\}$ separately. 
This implies the following constraints on the coefficients $\mathcal{C}^{(n)}_{\ell} $,
\begin{itemize}
  \item (i) Under $i_1 \leftrightarrow i_2 $, $i_3 \leftrightarrow i_4$ exchange, $\mathcal{C}^{(n)}_{\ell} $ is symmetric.
  \item (ii) For $i_1=i_4$ and $i_2=i_3$, $\mathcal{C}^{(n)}_{\ell} $ must be positive. 
  \item (iii) For $i_1\neq i_2\neq i_3\neq i_4$, the  $\mathcal{C}^{(n)}_{\ell} $s satisfy the following quadratic Schwarz inequality:
  \eq\label{Constraints}
 \mathcal{C}^{(n)}_{\ell}(i_1, i_2,i_2,i_1)\mathcal{C}^{(n)}_{\ell}(i_3, i_4,i_4,i_3)- \left(\mathcal{C}^{(n)}_{\ell}(i_1, i_2,i_3,i_4)\right)^2 \geq 0\,.
  \eqe 
 Note that the equality is satisfied if the spin-$\ell$ state is unique, or if there is more than one but their couplings are identical and thus represent a degeneracy. 
\end{itemize}
Because $\mathcal{C}^{(n)}_{\ell} $ is a function of the $v_i$, the above conditions are now translated into constraints on the function $f_p (z)$. In other words, if the integral formula in (\ref{MainAnsatz1}) is to yield a consistent space-time S-matrix, the power series of $f_p (z)$ must satisfy an infinite series of constraints!

In the following, we will demonstrate that condition (i) implies each $f_p(z)$ must be the global conformal block\footnote{Note that there is an extra factor of $z^{{-}i_3{-}i_4}$ compared to the usual definition of global blocks. This factor is associated with the prefactors of the four-function which are partially canceled by the Koba-Nielsen factor. This will be discussed in detail shortly. }
\eq
f_p(z)=\,z^{p{-}i_3{-}i_4}{}_2F_1(p{+}(i_2{-}i_1),p{+}(i_3{-}i_4),2p,z)\,,
\eqe
while (ii) and (iii) further require that the global blocks that differ by integer dimensions must be combined into further subsets:
\eq
f_p(z)+\sum_{q=1} \chi_q f_{p{+}q}(z)\,,
\eqe
with the Virasoro blocks living at the ``boundary" of this subset. 
\subsection{Global blocks from exchange symmetry}
The exchanging symmetry of $\mathcal{C}^{(n)}_{\ell}$ under  $i_1 \leftrightarrow i_2 $, $i_3 \leftrightarrow i_4$  imposes constraints on the form of $v_i$, as seen in eq.(\ref{Cexamp}-\ref{Cexampf}). We will begin by assuming that $v_n$ takes the following factorized form:
\eq
v_n=\frac{\mathcal{F}(i_1,i_2,p)\mathcal{F}(i_3,i_4,p)}{\mathcal{G}(p)}\,.
\eqe
This form is motivated by the fact that it is associated with the product of two three-point functions. Note that the kinematic part of (\ref{AmplitudeFac}) suggests that for spin-$\ell$ exchange the residue is further symmetric under sole $i_1 \leftrightarrow i_2 $ exchange for $\ell\in even$ and anti-symmetric for $\ell\in odd$ (and similarly for $i_3 \leftrightarrow i_4 $). However, this conclusion is too hasty, as the coupling constants can also introduce compensating transformation properties, as in the case of structure constants of non-abelian algebra $f^{abc}$. In light of this, we will only require that the residue has definite parity under the combined exchange  $i_1 \leftrightarrow i_2 $ and $i_3 \leftrightarrow i_4 $ .

Let us begin with $v_1$, which appears by itself in the level-1 scalar coefficient $\mathcal{C}^{(1)}_{0}$ in eq.(\ref{Cexamp}) and level-2 spin-1 coefficient  $\mathcal{C}^{(2)}_{1}$ in eq.(\ref{Cexamp2}). Beginning with the ansatz 
\eq
v_{1}=\frac{(a_1i+a_2i_2+a_3p+a_4i_1^2+a_5i_2^2+a_6i_1i_2)(a_1i_4+a_2i_3+a_3p+a_4i_4^2+a_5i_3^2+a_6i_3i_4)}{a_7+a_8 p+a_9 p^2}\,,
\eqe
by simply solving 
\eq
\mathcal{C}^{(1)}_{0}(i_1,i_2,i_3,i_4)-\mathcal{C}^{(1)}_{0}(i_2,i_1,i_4,i_3)=\mathcal{C}^{(2)}_{1}(i_1,i_2,i_3,i_4)-\mathcal{C}^{(2)}_{1}(i_2,i_1,i_4,i_3)=0\,,
\eqe
and requiring definite parity, we find a unique solution:
\eq
\label{gb1}
v_1=\frac{(-i_1+i_2+p)(i_3-i_4+p)}{2 p}\,.
\eqe
Note that this leads to $\mathcal{C}^{(1)}_{0}=0$ and  
\eq
\mathcal{C}_1^{(2)}=-\frac{(i_1{-} i_2) (i_3 {-} i_4)}{4 (d{-}3) p (1{+}p)}\,,
\eqe
i.e. $\mathcal{C}_1^{(2)}$ is antisymmetric under  $i_1 \leftrightarrow i_2 $, $i_3 \leftrightarrow i_4$ exchange respectively. We will come back to this property shortly.

Moving on to $v_2$, we begin with the ansatz
\eqa
v_2&=&\frac{\mathcal{F}(i_1,i_2,p)\mathcal{F}(i_3,i_4,p)}{(b_{1}+b_{2}p+b_{3}p^2)}\, ,\\
\mathcal{F}(i_1,i_2,p)&=&a_1 i_1^2 {+}a_2i_2^2{+}a_3p^2{+}a_4i_1i_2{+}a_5i_1p{+}a_6i_2p{+}a_7i_1{+}a_8i_2{+}a_9p{+}a_{10}\,.
\eqae
Equipped with $v_1$, once again solving 
\eq
\mathcal{C}^{(2)}_{0}(i_1,i_2,i_3,i_4)-\mathcal{C}^{(2)}_{0}(i_2,i_1,i_4,i_3)=0\,,
\eqe
while requiring that $\mathcal{C}_0^{(2)}$ has definite parity under individual exchange leads to 
\eq
\label{gb2}
v_2=\frac{(-i_1+i_2+p) (-i_1+i_2+p+1) (i_3-i_4+p) (i_3-i_4+p+1)}{4 p (2 p+1)}\,.
\eqe
This pattern continues, at each level-$n$, the symmetric property of the scalar coefficient $\mathcal{C}_0^{(n)}$ algebraically leads to a unique solution for $v_n$. For example at level-3 we find:
\eq
\label{gb3}
v_3=-\frac{(i_1{-}i_2{-}p{-}2) (i_1{-}i_2{-}p{-}1) (i_1{-}i_2{-}p) (i_3{-}i_4{+}p) (i_3{-}i_4{+}p{+}1) (i_3{-}i_4{+}p{+}2)}{24 p (p{+}1) (2 p{+}1)}.
\eqe

It is straightforward to see that the expressions (\ref{gb1}), (\ref{gb2}) and (\ref{gb3}) precisely match the $z$ expansion of the global conformal block \cite{Perlmutter:2015iya}:
\eqa \label{confblock}
&&z^{p{-}i_3{-}i_4}{}_2F_1(p{+}(i_2{-}i_1),p{+}(i_3{-}i_4),2p,z)=z^{p{-}i_3{-}i_4}\left(1{+}\frac{({-}i_1{+}i_2{+}p)(i_3{-}i_4{+}p)}{2p}z \right.\\
\nonumber&&{+}\frac{({-}1{+}i_2{+}p)(1{-}i_1{+}i_2{+}p)(i_3{-}i_4{+}p)(1{+}i_3{-}i_4{+}p)}{4p(1{+}2p)}z^2\\
\nonumber&&\left.{+}\frac{({{-}}i_1{{+}}i_2{{+}}p) ({{-}}i_1{{+}}i_2{{+}}p{{+}}1) ({{-}}i_1{{+}}i_2{{+}}p{{+}}2) (i_3{{-}}i_4{{+}}p) (i_3{{-}}i_4{{+}}p{{+}}1) (i_3{{-}}i_4{{+}}p{{+}}2)}{24 p (p{{+}}1) (2 p{{+}}1)}z^3+\cdots\right)\,.
\eqae
Thus we see that when interpreted as a space-time scattering amplitude, exchange symmetry of the residue directly leads to $f(z)$ in (\ref{MainAnsatz2}) to be given by a sum over SL(2,R) global conformal blocks!

As noted previously, the conformal block above is off by a factor of $z^{{-}i_3{-}i_4}$ compared with the usual definition. This can be traced back to the original form prior to SL(2,R) gauge fixing:
\eq
A(s,t) =\int_0^1 dz_3\;  \prod_{i<j} z^{2k_i\cdot k_j}_{ij} \frac{\left(\frac{z_{14}}{z_{24}}\right)^{i_2{-}i_1} \left(\frac{z_{14}}{z_{13}}\right)^{i_3{-}i_4}}{z^{i_1{+}i_2}_{12}z^{i_3{+}i_4}_{34}} \sum_{p} C_{p} z^p{}_2F_1(p{+}(i_2{-}i_1),p{+}(i_3{-}i_4),2p,z)\,.
\eqe
Taking $(z_1,z_2,z_3,z_4)=(\infty,1,z,0)$ and including the gauge fixing factor we find
\eq
(\infty)^{2(m_1^2-i_1+1)}\int_0^1 dz_3\;  z^{\alpha' 2k_3\cdot k_4}(1-z)^{\alpha' 2k_2\cdot k_3}  \sum_{p} C_{p} z^{p{-}i_3{-}i_4}{}_2F_1(p{+}(i_2{-}i_1),p{+}(i_3{-}i_4),2p,z)\,.
\eqe
The factor $\infty$ drops out if we have $m_1^2-i_1+1=0$, which is the expected relation between the mass and the SL(2,R) conformal dimension $m^2=h-1$.

\begin{figure}
\begin{center}
\includegraphics[scale=0.6]{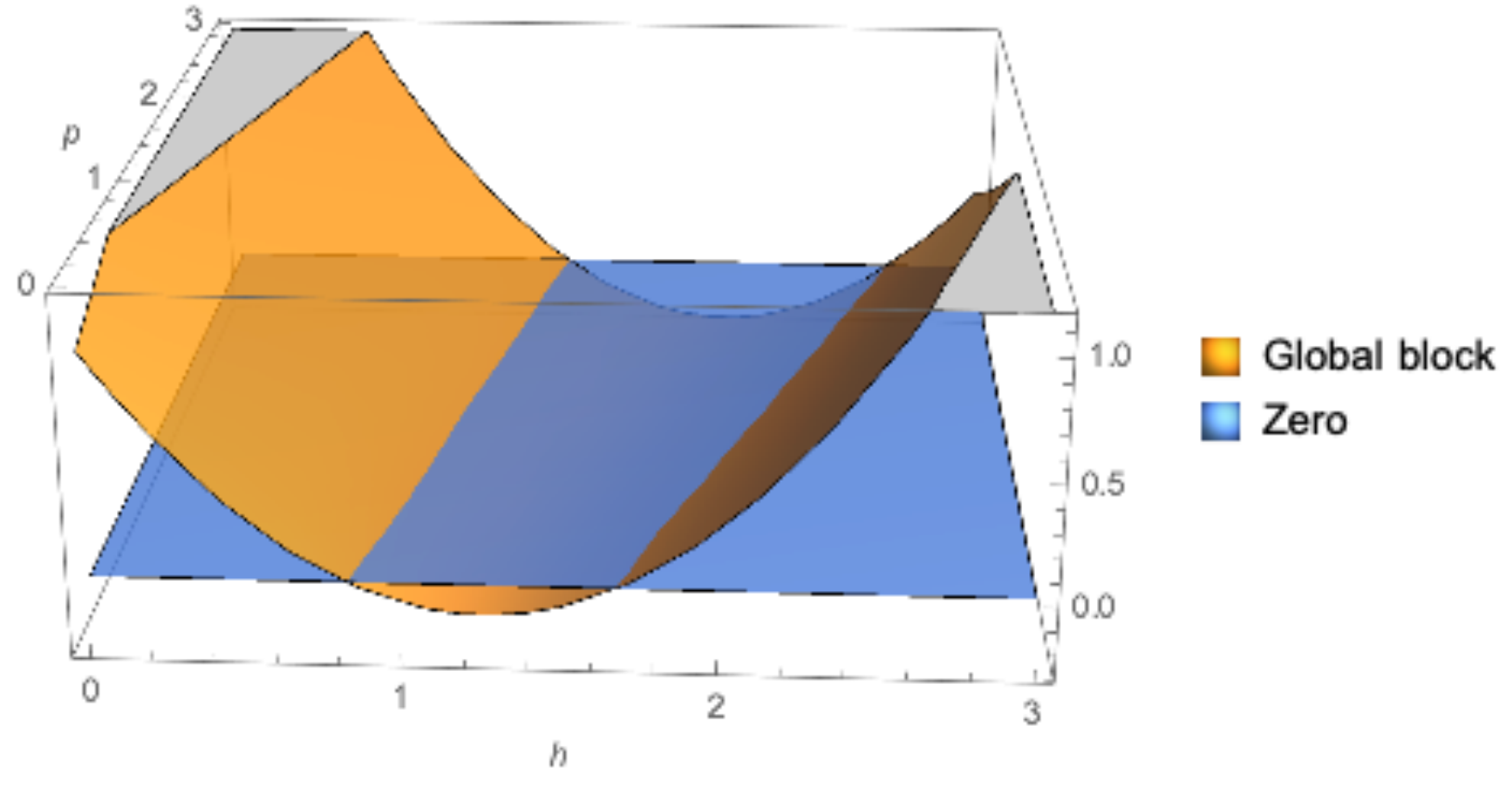}
\caption{The scalar coefficient $\mathcal{C}_0^{(2)}$ for the conformal block $(p,h)$. We see that there are regions of $(p,h)$ where the coefficient becomes negative, thus violating unitarity.}
\label{PositivityPlot}
\end{center}
\end{figure}

We have concluded that if the residue is to respect exchange symmetry, the ``dilatation blocks" must be linearly combined into an SL(2,R) global conformal block. We now move on to the factorization constraints (ii) and (iii) in (\ref{Constraints}). Let us consider the simplest condition where $i_1=i_2=i_3=i_4=h$. We display the four dimensional scalar coefficient at level 2, $\mathcal{C}_0^{(2)}$, as a function of $(p,h)$ in Figure (\ref{PositivityPlot}). We can see that there are regions of conformal dimension both for the external ($h$) and internal ($p$) that violate positivity bounds. In general, at level $n$, the spin $n$ coefficient derived from the global block will automatically satisfy the factorization constraints (ii) and (iii) in (\ref{Constraints}), but cease to do so when $\ell<n$. This suggests that one must further consider linear combinations of global blocks, which will be the focus in the next section.

\subsection{Non-negativity and the Virasoro block}
We have seen that given a global block, for identical external states the residue coefficients can easily violate positivity. To remedy the situation, we can consider linear combinations of global blocks. Note that since for any global block of conformal dimension $p$, the $s$-channel singularity occurs at $s=n{+}p{-}1$ with $n\in$ integer, we must consider linear combinations of blocks that differ by integer numbers. Thus we will be considering the following linear combination:
\eq\label{combconfblock}
f_p(z)=\sum_{a=0} \chi_a z^{p-i_3-i_4+a}{}_2F_1((p+a)+(i_2-i_1),(p+a)+(i_3-i_4),2(p+a),z)\,,
\eqe
where $\chi_0=1$ and in general $\chi_a$ can be a function of $\{i_1,i_2,i_3,i_4,p,d\}$. Note that as the coefficients $\mathcal{C}_\ell^{(n)}$ are $d$ dependent, their positivity will translate to the  $d$ dependence for $\chi_a$. Taking linear combinations of global blocks has the interpretation of taking linear combinations  of SL(2,R) primaries, and thus the coefficients $\chi_a$ should once again take on factorized form reflecting their OPE nature. We therefore introduce an ansatz for $\chi_a$ of the form:
\eq
\chi_a=\frac{\mathcal{F}(i_1,i_2,p)\mathcal{F}(i_4,i_3,p)}{\mathcal{G}(p)}\,.
\eqe
We will be looking for polynomial solutions to $\mathcal{F}(i,j,p)$ that are of lowest degree in $i,j$.  Next, we impose the factorization constraints in (\ref{Constraints}) (ii), (iii)
\eqa
&&(ii)\quad \mathcal{C}_\ell^{(n)}(i_1,i_2,i_2,i_1)>0  \label{Constraints2a} \,,\\
&&(iii)\quad \mathcal{C}_\ell^{(n)}(i_1,i_2,i_2,i_1)\mathcal{C}_\ell^{(n)}(i_3,i_4,i_4,i_3)-(\mathcal{C}_\ell^{(n)}(i_1,i_2,i_3,i_4))^2\geq0\,,  \label{Constraints2b}
\eqae
which will carve out the solution space for $\chi_a$.  Remarkably, we will find that the Virasoro block lives on the boundary of the solution space!

Let us analyze the constraints one level at a time.

\noindent\textbf{Level one:}\\
Up to level one, we have spin-0 and spin-1 coefficients $(\mathcal{C}_{0}^{(1)}, \mathcal{C}_{1}^{(1)})$, and only the scalar coefficient depends on $\chi_1$, given as:
\eq
\mathcal{C}_0^{(1)}=\chi_1(i_1,i_2,i_3,i_4,p)\, .
\eqe
Now we ask if the scalar coefficient can saturate the bound in (\ref{Constraints2b}), for all positive external dimensions $(i_1, i_2, i_3, i_4)$.  We find that a large family of solutions exists, of the form: 
\eq
\chi_1=(i_1-i_2)(i_4-i_3)\mathcal{F}(p)\,.
\eqe
The solution is minimal in the sense that it is lowest degree in external dimensions. With the above, (\ref{Constraints2a}) becomes:
\eqa
\mathcal{C}_0^{(1)}(i_1,i_2,i_2,i_1)=(i_1-i_2)^2\mathcal{F}(p)\geq0\,.
\eqae
Thus the ``boundary" of  (\ref{Constraints2a})  and (\ref{Constraints2b}) corresponds to $\mathcal{F}(p)=0=\chi_1$.

\noindent\textbf{Level two:}\\
At level two, while we have spins up to 2, $\chi_2$ only appears in the scalar coefficient $\mathcal{C}_0^{(2)}$. We will set $\chi_1=0$ and write down a minimal ansatz for $\chi_2$:
\eq\label{chi2Ansatz}
\chi_2=\frac{(a_1 i_1{+}a_2 i_2{+}a_3 i_1^2{+}a_4 i_2^2{+}a_5 i_1 i_2{+}a_6p{+}a_7i_1p{+}a_8i_2p{+}a_9p^2)(i_1\rightarrow i_4, i_2\rightarrow i_3)}{(b_0{+}b_1d{+}b_2 p{+}b_3dp{+}b_4p^2{+}b_5dp^2{+}b_6p^3)}\,.
\eqe
Note that we have allowed the dependence on $i_1, i_2,i_3,i_4$ and $p$ up to at least degree 2 in the numerator, as there are no solutions to (\ref{Constraints2a}, \ref{Constraints2b}) with lower degrees. We find that simply requiring the scalar coefficient to satisfy the equality in (\ref{Constraints2b}) for all external dimensions,
\eq
\mathcal{C}_{0}^{(2)}(i_1,i_2,i_2,i_1)\mathcal{C}_{0}^{(2)}(i_3,i_4,i_4,i_3)-(\mathcal{C}_{0}^{(2)}(i_1,i_2,i_3,i_4))^2=0\,,\quad \forall i_1, i_2,i_3,i_4\geq 0
\eqe
while respecting (\ref{Constraints2a}), the $\chi_2$ ansatz in (\ref{chi2Ansatz}) is reduced to
\eq\label{chi2}
\chi_2=\frac{(3 i_1^2 {+} 3 i_2^2 {-} i_2 (1{+}6 a p){-}i_1 (1{+} 6 a p){-}6i_1i_2{+} p ({-}5 {-} 3 p {+} 6 x (3 {+} p)))(i_1\rightarrow i_4, i_2\rightarrow i_3)}{(2d (1 {+} 6ap)^2{-}4 (1{+}p) (13{-}6 ({-}7{+}4 a) p{+}36 ({-}1{+}a)^2 p^2)))}\,,
\eqe
where the variable $a$ is the ratio $a=a_7/a_5$. This is constrained to $-1.25\leq a\leq1/3$ as we now see. Note that the coefficient $\mathcal{C}_{0}^{(2)}$, with $\chi_2$ given in (\ref{chi2}), becomes:
\eq
\mathcal{C}_{0}^{(2)}=\frac{\mathcal{N}(i_1,i_2,p,a)\mathcal{N}(i_4,i_3,p,a)}{8 ({-}1{+}d) (1{+}p)^2 (1{+}2 p) ({-}d (1{+}6pa)^2{+}2 (1{+}p) (13{-}6 ({-}7 {+} 4a) p {+} 36 ({-}1 {+}a)^2 p^2))}\,.
\eqe
If one requires the coefficient to be positive when $i_1=i_4$, $i_2=i_3$, then the denominator must be positive
\eq
{-}d (1{+}6pa)^2{+}2 (1{+}p) (13{-}6 ({-}7 {+} 4a) p {+} 36 (a{-}1)^2 p^2)\geq0\, .
\eqe
It is easy to see positivity for positive $p$ in dimensions below $d=26$ will bound $-1.25\leq a\leq1/3$. We can consider the positivity constraint at higher levels. For example, the positivity of $\mathcal{C}_{1}^{(3)}$ will further restrict the region to $0.32\leq a\leq1/3$. As the lower bound asymptotes to match with the upper bound, fixed at $1/3$,  being at the boundary of (\ref{Constraints2a}, \ref{Constraints2b}) uniquely determines $\chi_2$ to be:
\eq
\label{vb2}
\chi_2=\frac{({-}3 (i_1{-}i_2)^2{+}(i_1{+}i_2){+}2p(i_1{+}i_2){+}(p{-}1) p)({-}3 (i_4{-}i_3)^2{+}(i_4{+}i_3){+}2p(i_4{+}i_3){+}(p{-}1) p)}{2(1+2p) (26-d+42p-2dp+16p^2)}\,.
\eqe

\noindent\textbf{Level three:}\\
Let us now move to level three, where we again begin with the scalar coefficient $\mathcal{C}_{0}^{(3)}$. The minimal ansatz for $\chi_3$ is given as:
\eqa
\chi_3&=&-\frac{T(i_1,i_2)T(i_4,i_3)}{b_0+b_1p{+}b_2p^2{+}b_3p^3{+}b_4p^4{+}b_5p^5{+}b_6d{+}b_7p d{+}b_8p^2 d{+}b_9p^3d{+}b_{10}p^4 d}\nonumber\\
\nonumber T(i,j)&=&(a_1 i^2{+}a_2 j^2{+}a_3 i\, j{+}a_4 i^3{+}a_5 i^2\,j{+}a_6i\,j^2{+}a_7j^3{+}a_8i\,p\\
&&{+}a_9jp{+}a_{10}i^2p{+}a_{11}j^2p{+}a_{12}i j p+a_{13}ip^2{+}a_{14} jp^2)
\eqae
Setting $\chi_1=0$ and $\chi_2$ to (\ref{vb2}), once again by imposing (\ref{Constraints2a})  and equality in (\ref{Constraints2b}) for $\mathcal{C}_{0}^{(3)}$, 
\eq
\label{s3cf}
\mathcal{C}_{0}^{(3)}(i_1,i_2,i_2,i_1)\mathcal{C}_{0}^{(3)}(i_3,i_4,i_4,i_3)-(\mathcal{C}_{0}^{(3)}(i_1,i_2,i_3,i_4))^2=0, \quad \forall i_1, i_2,i_3,i_4\geq 0
\eqe
the ansatz for $\chi_3$ can be completely fixed up to the ratio $a=a_{11}/a_2$, which is confined to the region  $-6.58\leq a\leq1$. This range is considerably reduced by considering the positivity in (\ref{Constraints2a}) for $\mathcal{C}_1^{(4)}$, restricting to $0.95\leq a\leq1$. The positivity of $C_{n{-}3}^{(n)}$ in (\ref{Constraints2a}) at higher $n$ further pushes the lower bound to $1$, thus fixing all the ansatz to: 
\eq
\label{vb3}
\chi_3={-}\frac{(i_1{-}i_2) (i_3{-}i_4) ({-}i_1 {+} i_1^2 {-} i_2  {+} i_2^2{-} 2 i_1 i_2 {+} p {-} i_1 p {-} i_2 p) ({-}i_3 {+} i_3^2 {-} i_4 {+} i_4^2{-} 2 i_3 i_4  {+} p {-} i_3 p {-} i_4 p)}{2 p (p{+}1) (p{+}2) (28{-}d{+}19p{-}dp{+}3p^2)}\,.
\eqe

Before moving on to level $4$, let us compare ($\chi_1$, $\chi_2$, $\chi_3$) in (\ref{vb2}) and (\ref{vb3}) to the Virasoro block expansion on the global blocks \cite{Perlmutter:2015iya}: 
\eq
V_{i_1,i_2,i_3,i_4}=\sum_{a=0}^\infty\,  u_a z^{p-i_3-i_4+a}{}_2F_1((p+a)+(i_2-i_1),(p+a)+(i_3-i_4),2(p+a),z)\,,
\eqe
where 
\begin{align}
u_0=&1\nonumber\\
u_1=&0\nonumber\\
u_2=&\frac{(i_1 {-} 3 i_1^2 {+} i_2 {-} 3 i_2^2{+} 6 i_1 i_2  {-} p {+} 2 i_1 p {+} 2 i_2 p {+} p^2) (i_3 {-} 3 i_3^2 {+} i_4  {-} 3 i_4^2{+} 6 i_3 i_4 {-} p {+} 2 i_3 p {+} 2 i_4 p {+} p^2)}{2 (1 {+} 2 p) (c {+} 2 c p {+} 2 p ({-}5 {+} 8 p))}\nonumber\\
u_3=&{-}\frac{(i_1{-}i_2) (i_3{-}i_4) ({-}i_1 {+} i_1^2 {-} i_2  {+} i_2^2{-} 2 i_1 i_2 {+} p {-} i_1 p {-} i_2 p) ({-}i_3 {+} i_3^2 {-} i_4 {+} i_4^2{-} 2 i_3 i_4  {+} p {-} i_3 p {-} i_4 p)}{2 p (p{+}1) (p{+}2) (c p{+}c{+}3 p^2{-}7 p{+}2)}\, .
\end{align}
We find that $\chi_i=u_i$ if we set the central charge to $c=26{-}d$! Thus we see that the Virasoro blocks sits at the boundary of the constraints in (\ref{Constraints2a}, \ref{Constraints2b}).

\noindent\textbf{Level Four:}\\
At level 4 a new phenomenon occurs. We find that by setting $\chi_i=u_i$ with $i=1,2,3$, there are no solutions for $\chi_4$ for which the equality in (\ref{Constraints2b}) when applied to $\mathcal{C}_{0}^{(4)}$ is saturated for all positive external dimensions. To see this, let us simplify the problem to the vacuum block, $p=0$, and set $i_1=i_2=i$ $i_3=i_4=j$. In this case a general ansatz for $\chi_4$ is given as:  
\eq\label{chi4a}
\chi_4=\frac{g(i)g(j)}{a_1 d^2 +a_2 d+a_3}\,.
\eqe
where $g(i)$ is some polynomial. The boundary of the constraint in (\ref{Constraints2b}) becomes
\eqa\label{equal1}
\mathcal{C}_{0}^{(4)}(i,i,i,i)\mathcal{C}_{0}^{(4)}(j,j,j,j)-(\mathcal{C}_{0}^{(4)}(i,i, j, j))^2=0\,,  \quad \forall i, j\geq 0\,.
\eqae
With (\ref{chi4a}) the scalar coefficient takes the form 
\eqa
\nonumber\mathcal{C}_0^{(4)}&=&\frac{49 (i^2{+}j^2)}{8 (d^2{-}1)}{+}\frac{(7 (5 d{-}142)) (i{+}j)}{48 (d^2{-}1)}{+}\frac{d (9 d{-}490){+}6704}{384 (d^2{-}1)}{+}\frac{189 i j (i{+}j)}{d^3{-}26 d^2{-}d{+}26}\\
&&{-}\frac{(2 (d{+}28)) i^2 j^2}{d^3{-}26 d^2{-}d{+}26}{+}\frac{(d (d{+}680){-}19166) i j}{30 (d{-}26) (d{-}1) (d{+}1)}+\frac{g(i)g(j)}{a_1d^2{+}a_2 d{+}a_3}\,.
\eqae
It is straightforward to see that there are no polynomial solutions for $g(i)$ that satisfy (\ref{equal1}). Thus there are no linear combinations of the global blocks that can saturate (\ref{Constraints2b}), so the true boundary is no longer given by the equality. 

To seek the boundary, we begin with the following minimal ansatz:   
\eq
\chi_4=\frac{(i+a_4 i^2) (j+a_4 j^2)}{a_1 d^2 +a_2 d+a_3}\,.
\eqe
For Virasoro block these parameters would be:
\eq
a_4=5, \; a_1=\frac{25}{2},\; a_2=-705,\; a_3=9880\,. 
\eqe
Now (\ref{equal1}) becomes:
\eqa\label{equal2}
&&\mathcal{C}_{0}^{(4)}(i,i,i,i)\mathcal{C}_{0}^{(4)}(j,j,j,j)-(\mathcal{C}_{0}^{(4)}(i,i, j, j))^2=\nonumber\\
\label{inc4}&=&\frac{(i{-}j)^2}{(d{-}26)^2(d{-}1)^2(d{+}1)}\left(f_0(j,d){+}f_1(j,d)i{+}f_2(j,d)i^2\right)>0\,,  \quad \forall i, j\geq 0
\eqae
where $f_i(j,d)$'s are polynomial functions $j$ with maximal degree 2. Since positivity rests on the sign of the last parenthesis, we can consider the space in $\{a_i\}$ carved out by the requirement that the polynomial inside can at most have a single root, so that it is never negative.\footnote{Of course for specific $i,j$, $\mathcal{C}_{0}^{(4)}(i,i,i,i)\mathcal{C}_{0}^{(4)}(j,j,j,j)-(\mathcal{C}_{0}^{(4)}(i,i, j, j))^2$ may vanish. The previous discussion with regards to the boundary is about vanishing for \textit{all} positive $(i,j)$.} For example in $d=14$, the discriminant of $i$ in the last parenthesis of (\ref{inc4}) is:
\eq
f_1(j,d)^2-4f_0(j,d)f_2(j,d)=(a_4-5)^2\mathcal{F}(j,a_1,a_2,a_3,a_4)\,.
\eqe
Importantly, for any value of $\{a_1,a_2,a_3,a_4\}$ the function $\mathcal{F}$ will always have regions in $j$ for which it is positive, implying the existence of solutions for which (\ref{equal2}) is negative.  This tells us that $a_4=5$.

\begin{figure}
\begin{center}
\includegraphics[scale=0.4]{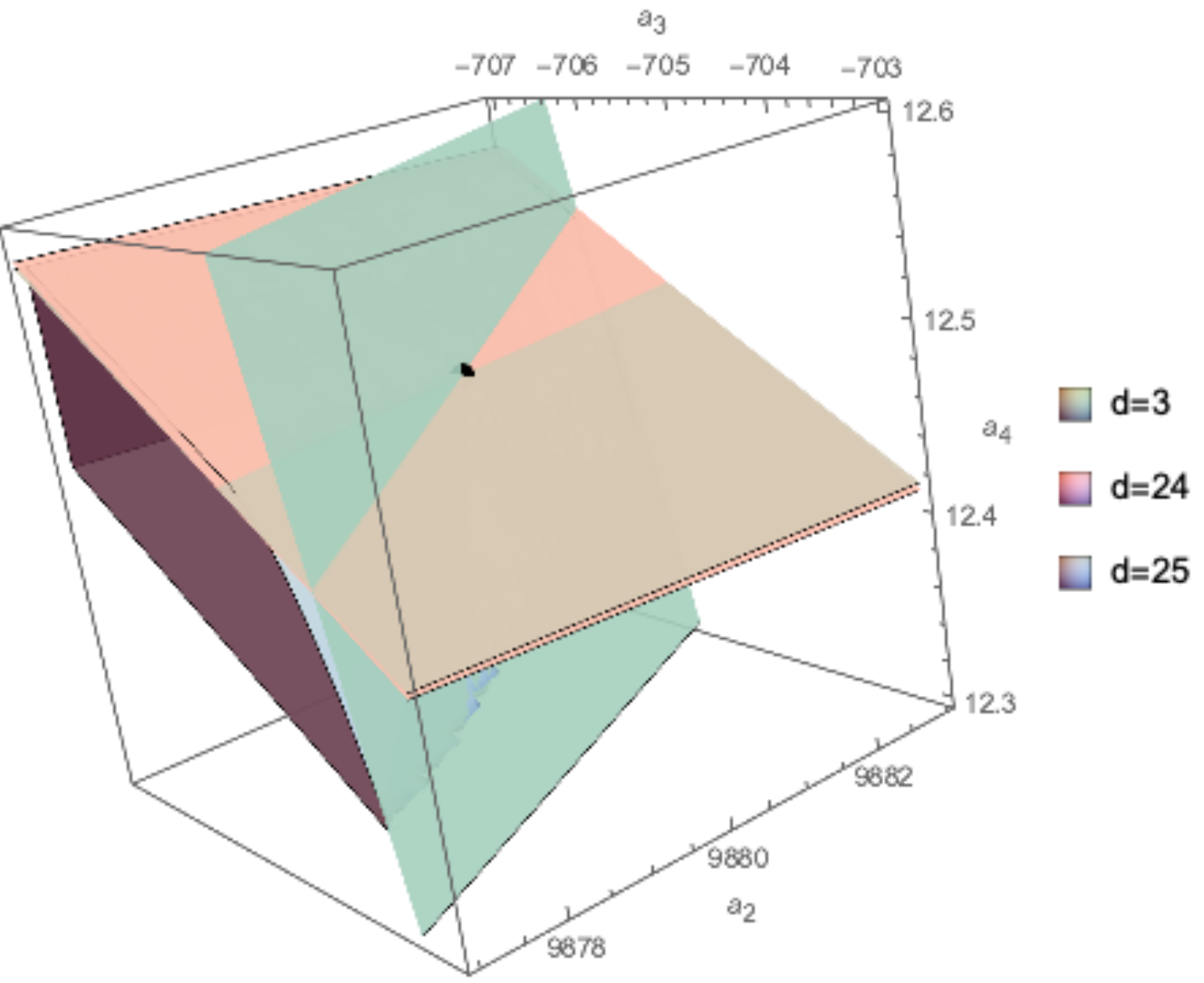}\quad\quad\includegraphics[scale=0.5]{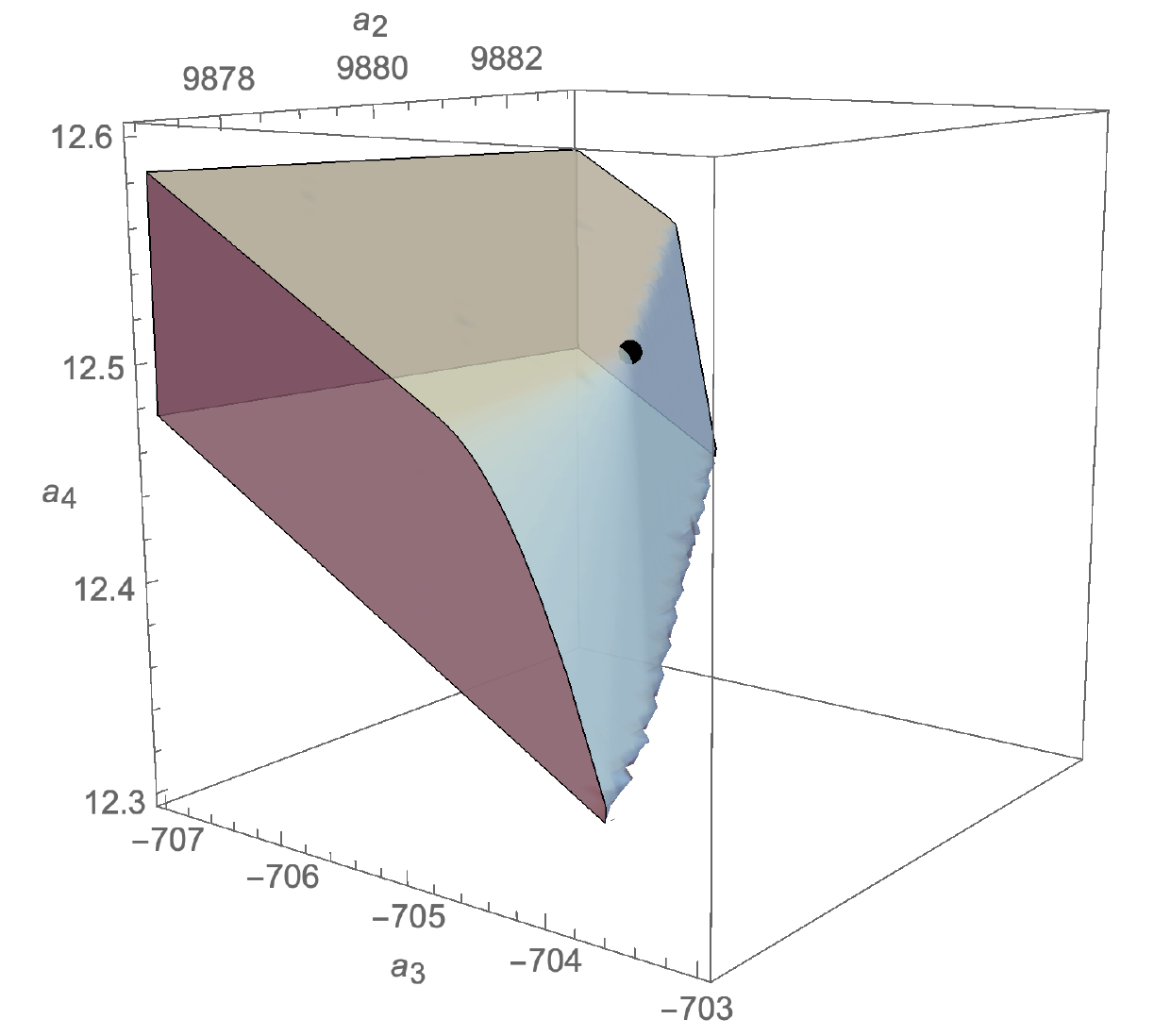}
  \caption{Solution space for $(a_1,a_2,a_3)$. This diagram is carved out by inequalities corresponding to different dimensions, with $d=3,24,25$ shown here. In each dimension, the saturated inequality is a two-dimensional plane. The Virasoro point lives on the intersection of these planes.}
 \label{3DPlot}
\end{center}
\end{figure}

\begin{figure}
  \begin{center}
\includegraphics[scale=0.4]{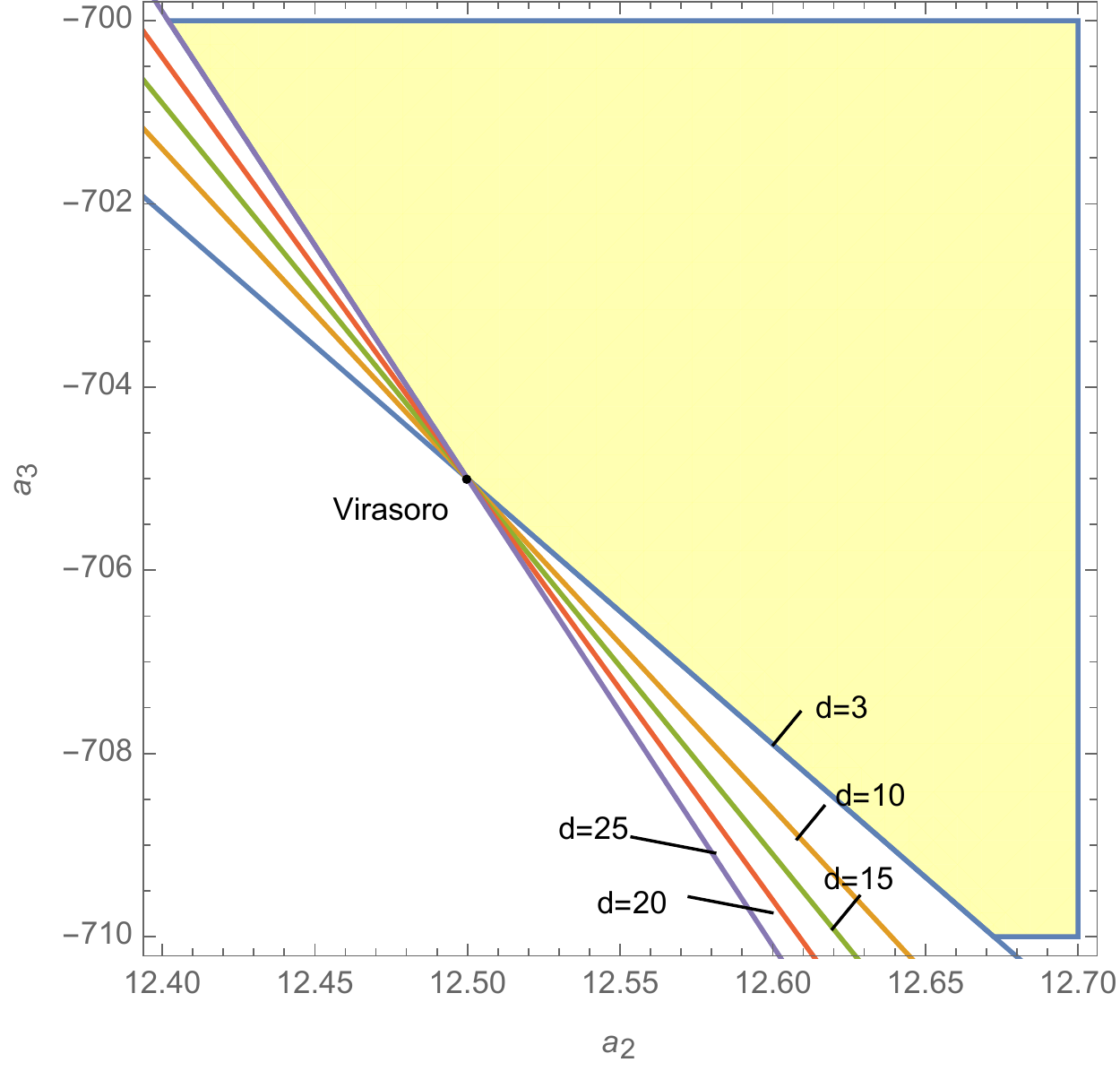}
        \caption{$a_1=9880$ solution space. The point in figure is the Virasoro coefficient.}
          \label{2DPlot}
     \end{center}
\end{figure}

For the remaining coefficients, lets us see the image of (\ref{equal2}) in $(a_1, a_2, a_3)$ space. With $a_4=5$,  (\ref{equal2}) yields the following inequalities: 
\eqa
d=3&:& 0<9a_1 + 3a_2+a_3 \leq 15755/2\,,\\
d=24&:&0 < 576a_1 + 24 a_2 + a_3 \leq 160\,, \\
d=25&:&0 < 625a_1 + 25a_2+ a_3 \leq 135/2\,.
\eqae
The carved out region is displayed in Figure (\ref{3DPlot}), where we have displayed the region as well as the respective hyperplane implied by the above inequalities. One can see from the figure that the Virasoro coefficients are set at a kink in the boundary. To make the last property manifest, let us consider a two-dimension sub-plane in Figure (\ref{3DPlot}) defined by  $a_1=9880$, as shown in Figure (\ref{2DPlot}). Then the Virasoro point is at the kink defined by the intersection of the constraints $d=3$ and $25$. 

In summary, by considering the boundary carved out by the factorization conditions in (\ref{Constraints2a}) and (\ref{Constraints2b}), we find that the Virasoro block sits at unique special points on these boundaries.

\section{Review of the EFThedron}\label{sEFT}
The unitarity constraint on factorization also leaves its fingerprint on the low energy amplitude, in the form of positivity bounds on the EFT couplings. A precursor of such bounds is the positivity of the leading four derivative operator derived from optical theorem in~\cite{Adams:2006sv}. More recently, an infinite set of positivity bounds have been derived by considering the near forward limit of the low energy expansion, exploiting the positive geometry that arises from expanding the Gegenbauer polynomials~\cite{EFThedron}. Here we present a brief review.

Let us consider the low energy limit of an UV complete amplitude. Here, by low energy we are referring to the limit where the Mandlestam variables are smaller than the scale set by the UV massive states. For an ordered amplitude, in this limit the amplitude takes the form:
\eq
A(s,t)|_{s,t\ll 1}=\frac{a}{st}+\frac{b_1}{s}{+}\frac{b_2}{t}+\sum_{k,q\geq 0} g_{k,q}s^{k{-}q}t^q\,,
\eqe 
where we have set the UV scale to $1$, and $a, b_1, b_2$ can be some kinematic dependent functions. Cyclic symmetry means that we can identify $g_{k,q}=g_{k,k{-}q}$. Note that we have defined the EFT couplings $g_{k,q}$ from the polynomial expansion of the low energy amplitude. This allows us to define the couplings via a contour integral in the complex $s$-plane, with $t$ held fixed:
\eq\label{gder}
g_{k,q}=\frac{1}{q!}\frac{d^q}{dt^q}\left(\frac{i}{2p}\oint\frac{ds}{s^n} A(s,t)\right)\bigg|_{t=0}\,.
\eqe
By deforming the contour, one picks up the residues and discontinuity on the real positive $s$-axes (since we do not have $u$-channel thresholds), arising from  
\eq\label{threshold}
A(s,t)|_{s\rightarrow m^2}=\sum_{\ell} c_\ell\frac{G_{\ell}^{d}(\cos\theta)}{s-m^2},\quad \cos\theta=1{+}\frac{2t}{m^2} \,,
\eqe
where $c_\ell\geq0$. While the above form describes the behavior of the amplitude near tree-threshold, near the forward limit the same form holds for branch cuts, except that one has to sum over a continuous spectrum.

In other words, combining (\ref{gder}) with (\ref{threshold}), we have that the low energy coupling can be matched to the derivative expansion of the Gegenbauer polynomials as well as the propagators. Defining the Taylor coefficients $v_{\ell,q}$ from:
\eq
G_\ell^{d}(1+\delta)=\sum_{q=0}v_{\ell,q}\delta^q\,,
\eqe
at fixed $k$, fixed mass dimension of the operator, we find:
\eq
\vec{g}_k=\sum \left(\begin{array}{c} g_{k,0} \\ g_{k,1} \\ \vdots \\ g_{k,n}\end{array}\right)\in\sum_{a} c_a \vec{\mathcal{G}}_{\ell_a},\quad 
 \vec{\mathcal{G}}_{\ell}=\left(\begin{array}{c} v_{\ell, 0} \\ v_{\ell, 1} \\ \vdots \\ v_{\ell, n}\end{array}\right)\,,
\eqe
where $a$ labels the spectrum of the UV states and $c_a>0$. The above implies the $n{+}1$ component vector $\vec{g}_k$ must lie inside the convex hull of the Gegenbauer vectors $\vec{\mathcal{G}}_{\ell}$. Importantly, due to the positivity properties of $\vec{\mathcal{G}}_{\ell}$, its convex hull is a cyclic polytope, and the boundary of the hull is constructed by adjacent pairs of $\vec{\mathcal{G}}_{\ell}$. For example for $n=4$,  $\vec{g}_k$ being inside the convex hull implies that:
\eq
\langle \vec{g}_k, \ell_i,\ell_i{+}1,\ell_j,\ell_j{+}1\rangle=\textrm{det}\left[\vec{g}_k, \vec{\mathcal{G}}_{\ell_i}, \vec{\mathcal{G}}_{\ell_i+1}, \vec{\mathcal{G}}_{\ell_j}, \vec{\mathcal{G}}_{\ell_j+1}\right]\equiv \vec{g}_k\cdot W_{I} \geq0\,,
\eqe
where we use $W_I$ as a short hand notation for the boundary, here given by $\langle *, \ell_i,\ell_i{+}1,\ell_j,\ell_j{+}1\rangle$. Note that since for any vector $\vec{g}_k$ that satisfies the above constraint, rescaling by a positive constant yields another solution, the geometry of the convex hull is really a polytope in $\mathbb{P}^n$.  For $n=odd$, we have the same pattern for the boundaries except with an extra vector  $\vec{\mathcal{G}}_{0}$.

We can also consider keeping $q$ fixed while collecting the couplings with distinct $k$. We then have:
\eq
\vec{g}_q=\sum \left(\begin{array}{c} g_{q,q} \\ g_{q+1,q} \\ \vdots \\ g_{q{+}n,q}\end{array}\right)\in\sum_{a} c_a v_q(x_a)^q\left(\begin{array}{c} 1 \\ x_a \\ \vdots \\ x^n_a\end{array}\right)\quad x_a=\frac{1}{m^2_a}\,.
\eqe
That is, the vector $\vec{g}_q$ lives in the convex hull of points on a moment curve, $(1,x,x^2,\cdots, x^n)$. This is a reflection that fixed $q$ means we are expanding the Gegenbauer polynomial to fixed order, and collecting the expansion of $1/(s-m^2)$ which gives a geometric series. Importantly, since we have  $x_a>0$, $\vec{g}_q$ really lives in the convex hull of a half moment curve $(1,x,x^2,\cdots, x^n)$ with $x\in R^+$. This implies that the Hankel matrix for the couplings, defined as the following symmetric matrix:
\eq
H=\left(\begin{array}{cccc} g_{k,q} & g_{k{+}1,q} & \cdots & g_{k{+}n,q} \\ g_{k{+}1,q} & g_{k{+}2,q} & \cdots & g_{k{+}1{+}n,q} \\ \vdots & \vdots & \ddots & \vdots \\ g_{k{+}1{+}n,q} & g_{k{+}2{+}n,q} & \cdots & g_{k{+}2n,q} \end{array}\right)\,,
\eqe
will have all non-negative minors.

Finally, the fact that the expansion in $s$ and $t$ corresponds to a direct product of geometries is reflected in the following product Hankel matrix constraint: 
\eq
H=\left(\begin{array}{cccc} \vec{g}_k\cdot W_{I}  & \vec{g}_{k+1}\cdot W_{I} & \cdots & \vec{g}_{k+1}\cdot W_{I} \\  \vec{g}_{k+1}\cdot W_{I} & \vec{g}_{k{+}2}\cdot W_{I} & \cdots &  \vec{g}_{k{+}1{+}n}\cdot W_{I} \\ \vdots & \vdots & \ddots & \vdots \\  \vec{g}_{k{+}1{+}n}\cdot W_{I} & \vec{g}_{k{+}2{+}n}\cdot W_{I} & \cdots &  \vec{g}_{k{+}2n}\cdot W_{I} \end{array}\right)\,,
\eqe
where $W_{I}$ is any one of the cyclic polytope boundaries in $\mathbb{P}^n$. 

The space of couplings carved out by the above inequalities is termed the EFThedron~\cite{EFThedron}, which we briefly summarize: 
\begin{itemize}
  \item Cyclic polytope constraints: 
  \eq
{\rm n\in even}: \langle \vec{g}_k, \ell_i,\ell_i{+}1,\ell_j,\ell_j{+}1,\cdots \rangle \geq0\,,\quad {\rm n\in odd}: \langle \mathbf{0}, \vec{g}_k, \ell_i,\ell_i{+}1,\ell_j,\ell_j{+}1,\cdots \rangle \geq0 \label{cpc}
\eqe
where $\mathbf{0}$ represents $\vec{\mathcal{G}}_0$.
  \item Hankel matrix constraints:
  \eq
 \textrm{Minor} \left[\begin{array}{cccc} g_{k,q} & g_{k{+}1,q} & \cdots & g_{k{+}n,q} \\ g_{k{+}1,q} & g_{k{+}2,q} & \cdots & g_{k{+}1{+}n,q} \\ \vdots & \vdots & \vdots & \vdots \\ g_{k{+}1{+}n,q} & g_{k{+}2{+}n,q} & \cdots & g_{k{+}2n,q} \end{array}\right]\geq0 \,,\label{hmc}
  \eqe
  \item Product Hankel matrix constraints:
  \eq
 \textrm{Minor} \left[\begin{array}{cccc} \vec{g}_k\cdot W_{I}  & \vec{g}_{k+1}\cdot W_{I} & \cdots & \vec{g}_{k+1}\cdot W_{I} \\  \vec{g}_{k+1}\cdot W_{I} & \vec{g}_{k{+}2}\cdot W_{I} & \cdots &  \vec{g}_{k{+}1{+}n}\cdot W_{I} \\ \vdots & \vdots & \vdots & \vdots \\  \vec{g}_{k{+}1{+}n}\cdot W_{I} & \vec{g}_{k{+}2{+}n}\cdot W_{I} & \cdots &  \vec{g}_{k{+}2n}\cdot W_{I} \end{array}\right]\geq0 \label{phc}\,.
  \eqe
\end{itemize}

\section{Implications of monodromy relations: the monodromy plane}\label{monosection}
The monodromy relation for  amplitudes of massless external states: 
\begin{equation}
\mathcal A\left(2134\right)+e^{i\pi  s}\mathcal A\left(1234\right)+e^{-i\pi u }\mathcal A\left(1324\right)=0\,,
\label{eq:nms}
\end{equation}
is a common feature shared by all flat-space open string amplitudes for identical external states. This relation reflects the fact that the corresponding worldsheet integrand is permutation invariant, and it is only the ordering of the integration regions that characterizes the distinct orderings of the amplitude.\footnote{For example the Lovelace-Shapiro amplitude: \[\mathcal A^{\left(LS\right)}\left(1234\right)=g^{2}\frac{\Gamma\left(\frac{1}{2}-s\right)\Gamma\left(\frac{1}{2}-t\right)}{\Gamma\left(-s-t\right)}\] will not satisfy the above monodromy relations since the vertex operators are not identical  \cite{Bianchi:2020cfc}} For general string compactifications, the function  $f\left(z\right)$ in  (\ref{IntroA})  can generate nontrivial monodromy around $z=0,1,\infty$. Assuming that the resulting monodromy is universal, i.e. it is simply an overall prefactor, the corresponding variation for \ref{eq:nms} will be 
\begin{equation}
\mathcal A\left(2134\right)+e^{i\pi  \left(s+a_s\right)}\mathcal A\left(1234\right)+e^{-i\pi  \left(u+a_u\right) }\mathcal A\left(1324\right)=0\,.
\label{eq:tms}
\end{equation}
Note that the monodromy relation, along with two other ones from permutations $1\leftrightarrow2$ and $1\leftrightarrow3$, result in 6 real conditions. Since we only have three independent amplitudes, $A\left(1234\right)$, $A\left(2134\right)$, and $A\left(1324\right)$,\footnote{With the cyclic invariance $A\left(1234\right)=A\left(4123\right)$ and reflection symmety $A\left(1234\right)=A\left(4321\right)$ any four-point amplitude with arbitrary order of $1,2,3,4$ can be identified with one of the three} in order for there to be a non-trivial solution we can only have $a_s=a_t=a_u= 0,\frac{2}{3},\frac{4}{3}$. To see this, we note that for there to be a solution the $6$  linear constrains from $3$ different monodromy relations must degenerate to at most $2$. Therefore there can be only one independent monodromy relation. In other words, the three different monodromy relations must be proportional to each other and this proportionality requirement fixes $a_s=a_t=a_u= 0,\frac{2}{3},\frac{4}{3} $.

It is instructive to see how such deformed monodromy arrises upon compactification of flat space string amplitudes. Consider the compactification of the Tachyon amplitude in bosonic string theory, with the 26 dimensional momenta denoted as
$K_{i}$ (with $\alpha'K_{i}^{2}=1$).
\begin{equation}
A^{\left(26D\right)}\left(1234\right)=\frac{1}{\text{Vol}}\int\prod_{i=1}^{4}dz_{i}\prod_{j<i=2}^{4}\left(z_{i}-z_{j}\right)^{2\alpha'K_{i}\cdot K_{j}}\label{eq:4bta}\,.
\end{equation}
By decomposing the momenta into $d$ and $26-d$ components, $K_{i}=\left(k_{i},q_{i}\right)$ with $k_{i}^{2}=0$, we obtain the compactified $d$-dimensional amplitude as:
\begin{equation}
A^{\left(4D\right)}\left(1234\right)=\frac{1}{\text{Vol}}\int\prod_{i=1}^{4}dz_{i}\prod_{j<i=2}^{4}\left(\left(z_{i}-z_{j}\right)^{2\alpha'k_{i}\cdot k_{j}}\left(z_{i}-z_{j}\right)^{2\alpha'q_{i}\cdot q_{j}}\right)\label{eq:c4bta}\,.
\end{equation}
Note that the above is permutation invariant under $\left\{ z_{i},k_{i}\right\} \leftrightarrow\left\{ z_{j},k_{j}\right\}$ only if all $q_{i}\cdot q_{j}$'s are equal. The mass shell condition 
\begin{align}
2\alpha'\left(q_{1}\cdot q_{2}+q_{1}\cdot q_{3}+q_{1}\cdot q_{4}\right)=-2\alpha'\left(k_{1}^{2}-K_{1}^{2}\right)=2\,,
\end{align} 
then fixes $2\alpha'q_{i}\cdot q_{j}$ to $\frac{2}{3}$, which leads to an additional phase of $\frac{2}{3}\pi$ or $\frac{4}{3}\pi$. Such ``twisted monodromy relations'' (\ref{eq:tms}) result in amplitudes without massless poles. Indeed expanding in $\alpha'$, the leading order identity is:
\begin{align}
A_{\textrm{IR}}\left(2134\right)+e^{2i\pi/3}A_{\textrm{IR}}\left(1234\right)+e^{4i\pi/3}A_{\textrm{IR}}\left(1324\right)=0\,,
\end{align}
which indicates that the leading order amplitude $A_{IR}$ can only be a constant. While such amplitudes are potentially interesting objects to study, we will be focusing on four-point amplitudes where massless poles are present, thus we will restrict ourselves to the standard monodromy relation (\ref{eq:nms}). Generalizing to the twisted case is straightforward.



 \subsection{Single color ordered amplitudes}
The monodromy relation (\ref{eq:nms}) imposes restrictive constraints on the amplitude. First of all, it implies integer spectrum of the theory, which is evident  by
looking at the imaginary part of the four-point monodromy relation $\sin\left(\pi s\right)\mathcal A\left(1234\right)=\sin\left(\pi u\right)\mathcal A\left(1324\right)$
 The $s$-channel  and  $u$-channel poles must be paired with the zeros in the respective sine factors for this identity
to hold for all values of $s,t,u$, while all zeros of the sine function are located at integer values. More importantly, it implies nontrivial mixing relations of couplings at different derivative order, which will be the focus of this subsection. 

We will be interested in the case where the complete amplitude is proportional to the leading order in $\alpha'$ expansion, 
\begin{equation}\mathcal{A}\left(1234\right)=S\left(\left\{ k_{i},\epsilon_{i}\right\} \right)A\left(s,t\right)=S\left(\left\{ k_{i},\epsilon_{i}\right\} \right)\left(-\frac{1}{st}+\cdots\right)\, ,
\label{eq:as}
\end{equation}
where $S$ is the universal common factor that is independent of the ordering. This occurs when we have four-dimensional massless external states, where the prefactor is completely fixed by the helicity weights, or maximal supersymmetry in general dimensions, where the external states are in the same multiplet. The monodromy relation (\ref{eq:nms}) then becomes 
\begin{equation}
A\left(s,u\right)+e^{i\pi  s}A\left(s,t\right)+e^{-i\pi u }A\left(t,u\right)=0\,,
\label{eq:ms}
\end{equation}
and the constraint at each derivative order can be explicitly written in terms of Laurent coefficients of $A\left(s,t\right)$. To study the constraint we first write the factor $A\left(s,t\right)$ in \ref{eq:as} as an expansion  
  \begin{equation}
  A\left(s,t\right)=-\frac{1}{st}+\left(\frac{b}{s}+\frac{b}{t}\right)+\left(c \frac{t}{s}+c\frac{s}{t}\right)+g_{00}+\left(g_{1,0}s+g_{1,1}t\right)+\sum_{k\geq q\geq0}g_{k,q}s^{k-q}t^{q}\label{eq:ALE}\, ,
  \end{equation} 
then solve for the monodromy constraints imposed by (\ref{eq:ms}) order by order in $\alpha'$. First, one immediately finds the coefficient for $t^{2n}/s$ must be zero, including $b=0$. For $c$, as we will demonstrate in the next section, the unitarity constraint will also set it to zero. Thus for the remainder of the discussion we will solve the monodromy relations with respect to 
\begin{equation}
  A\left(s,t\right)=-\frac{1}{st}+\sum_{k\geq q\geq0}g_{k,q}s^{k-q}t^{q}\label{eq:ALE2}\, .
  \end{equation} 
 As an example, the solution up to $k=4$ is given as: 
 \begin{equation}
\left(\begin{array}{ccccc}
g_{00}\\
g_{1,0} & g_{1,1}\\
g_{2,0} & g_{2,1} & g_{2,2}\\
g_{3,0} & g_{3,1} & g_{3,2} & g_{3,3}\\
g_{4,0} & g_{4,1} & g_{4,2} & g_{4.3} & g_{4,4}
\end{array}\right)=\left(\begin{array}{ccccc}
\frac{\pi^2}{6}\\
g_{1,0} & g_{1,0}\\
\frac{\pi^4}{90}& \frac{\pi^{4}}{360} &\frac{\pi^4}{90}\\
g_{3,0} &2 g_{3,0}-\frac{ \pi ^2 }{6}g_{1,0}&2 g_{3,0}-\frac{ \pi ^2 }{6}g_{1,0} & g_{3,0}\\
\frac{\pi^6}{945}& g_{4,1} & -\frac{\pi^{6}}{15120}+2g_{4,1} & g_{4,1} &\frac{\pi ^6}{945}
\end{array}\right)\label{eq:agkq}\,.
\end{equation}

We will refer to the above solution as the \textit{monodromy plane}, defining a subspace in the EFT couplings where the monodromy conditions are satisfied order by order.  Our next step will be to constrain the remaining parameters with positivity conditions. But first, let us compare our solution with the actual superstring $A\left(s,t\right)$ factor, given by \cite{Terasoma:1999}
\begin{equation}
A_{\textrm{Superstring}} \left(s,t\right)=\frac{\Gamma\left(-s\right)\Gamma\left(-t\right)}{\Gamma\left(1-s-t\right)}=\frac{1}{st}\exp\left(\sum_{n\geq2}\frac{\zeta\left(n\right)\left(\left(-s\right)^{n}+\left(-t\right)^{n}-\left(-s-t\right)^{n}\right)}{n}\right)\label{eq:str}\,.
\end{equation}
We can confirm that all fixed $g_{k,q}$ coefficients in $(\ref{eq:agkq})$  match with the string value after expressing the even zeta values, $\zeta\left(2\right)=\frac{\pi^2}{6}$, $\zeta\left(4\right)=\frac{\pi^4}{90}$, and so on. In particular, it will be useful later on to observe that the coefficients $g_{2n,0}$ are all set to $\zeta(2n)$ by monodromy.
But we also observe a correspondence between the remaining independent coefficients and the monomials of odd zeta values, a pattern that also persists to higher orders, as the string value of the free parameters on monodromy plane reads:
\begin{equation}
\label{fixzeta}
g_{1,0}=\zeta\left(3\right),\quad g_{3,0}=\zeta\left(5\right),\quad g_{4,1}=\frac{\pi^{6}}{1260}-\frac{1}{2}\zeta^{2}\left(3\right)\,.
\end{equation}
This is closely related to the fact that only certain linear combinations of the remaining free parameters survives double copy (we will show the explicit formula in \ref{eq:doublecopy}), and the closed superstring amplitude only contains odd zeta values \cite{Stieberger:2009rr,Schlotterer:2012ny}. 

This will become obvious when we analyze the monodromy relations order by order. First, let us write:
\begin{align}
A_u+e^{i \pi \alpha' s}A_s+e^{- i \pi \alpha' t }A_t=0\,,
\end{align}
where $A_u=A(s,u)$, $A_s=A(s,t)$, and $A_t=A(t,u)$, each with an expansion $A=A^0+\alpha' A^1+\alpha^2A^2+\ldots$. 
Expanding the monodromy relation, we obtain:

\begin{align}
(\alpha')^0:\ &A^0_s+A^0_t+A^0_u=0 \,,\nonumber \\
(\alpha')^1:\ &  A^1_s+A^1_t+A^1_u=0 \,, \nonumber \\
& s A_s^0-t A_t^0=0\,, \nonumber\\
(\alpha')^2:\ &  A^2_s+A^2_t+A^2_u-\frac{\pi^2}{2} (s^2A^0_s+t^2A^0_t)=0\,, \label{A2KK}\\
& s A_s^1-t A_t^1=0\,, \nonumber\\
(\alpha')^3:\ &  A^3_s+A^3_t+A^3_u-\frac{\pi^2}{2} (s^2A^1_s+t^2A^1_t)=0\,, \nonumber\\
& s A_s^2-t A_t^2-\frac{\pi^2}{6}(s^3A_s^0-t^3 A_t^0)=0\,, \label{A2BCJ}
\end{align}
we can easily recognize the expressions $A_s+A_t+A_u$ and $sA_s-t A_t$ as Kleiss-Kuijif (KK), and respectively Bern-Carrasco-Johansson (BCJ) relations \cite{Kleiss:1988ne,Bern:2008qj}, which will be relevant shortly. However, the above conditions are in general a form of modified KK and BCJ relations. For example, eqs.(\ref{A2KK}) and (\ref{A2BCJ}), relevant for the $A^2$ correction, read:
\begin{align}
&  \textrm{KK}[A^2]-\frac{\pi^2}{2} (s^2A^0_s+t^2A^0_t)=0\,, \\
& \textrm{BCJ}[A^2]-\frac{\pi^2}{6}(s^3A_s^0-t^3 A_t^0)=0\,.
\end{align}
It is clear that there will be at most two types of solutions for $A^2$ that satisfy the above equations: independent solutions that satisfy KK and BCJ relations, and solutions that satisfy the modified relations, and hence become entangled with lower mass dimension solutions, carrying relative factors of $\pi^2$. As shown in  \cite{Kawai:1985xq,Bern:2008qj,Stieberger:2009hq,BjerrumBohr:2009rd}, only KK and BCJ satisfying amplitudes are compatible with the double copy to closed strings. With the closed string containing only odd zeta values, such solutions must themselves correspond to an odd zeta value. This explains why we obtain independent families of solutions parameterized by even zeta values,  in particular why monodromy fixes $g_{2n,0}=\zeta(2n)$, while the coefficients not fixed by monodromy must contain odd zeta values.

As we will see, the EFThedron constraints will be able to fix the remaining parameters to (\ref{fixzeta}) with high precision.



\subsection{Bicolor ordered amplitudes }

Similarly we can study the monodromy constraint on theories with bicolor
structure: $\mathcal{A}\left(\left.{\cal P}\right|{\cal Q}\right)$,
where ${\cal P},{\cal Q}$ are two permutations of the color indices $\left\{ 1,2,3,4\right\} $. Z-theory provides an example of such amplitudes, which can be double-copied with Yang-Mills amplitudes to express the open superstring~\cite{Carrasco:2016ldy,Mafra:2016mcc}.

We will test whether the amplitude satisfies BCJ relations with respect to permutations of the set ${\cal P}$,
\begin{equation}
u\mathcal{A}\left(\left.1324\right|1234\right)=s\mathcal{A}\left(\left.1234\right|1234\right)\,,
\end{equation}
 and the monodromy relation with respect to the set ${\cal Q}$,
\begin{equation}
\mathcal{A}\left(\left.1234\right|2134\right)+e^{i \pi s}\mathcal{A}\left(\left.1234\right|1234\right)+e^{-i \pi u}\mathcal{A}\left(\left.1234\right|1324\right)=0\label{eq:monodromy}\,.
\end{equation}
This will help determine if the bicolor amplitude can be derived from a worldsheet amplitude of the following form
\begin{align}\nonumber
\mathcal{A}\left(\left.{\cal P}\right|{\cal Q}\right)
= & -S\left(\left\{ k_{i},\epsilon_{i}\right\} \right)\int_{0}^{1}dz_{{\cal P}\left(2\right)}f\left(z_i\right) z_{{\cal P}\left(2\right)}^{-2k_{{\cal P}\left(1\right)}\cdot k_{{\cal P}\left(2\right)}}\left(1-z_{{\cal P}\left(2\right)}\right)^{-2k_{{\cal P}\left(2\right)}\cdot k_{{\cal P}\left(3\right)}}\\
 & \times\frac{\left(z_{{\cal P}\left(1\right)}-z_{{\cal P}\left(3\right)}\right)\left(z_{{\cal P}\left(1\right)}-z_{{\cal P}\left(4\right)}\right)\left(z_{{\cal P}\left(3\right)}-z_{{\cal P}\left(4\right)}\right)}{\prod_{i=1}^{4}\left(z_{{\cal Q}\left(i\right)}-z_{{\cal Q}\left(i+1\right)}\right)}\,,
\end{align}
where $f\left(z_i\right) $ is a total symmetric function of all $z_i$'s, and has trivial monodromy.

For such bicolor amplitudes satisfying BCJ and monodromy relations there is only
one independent four-point amplitude, all other amplitudes with different
orders of the two sets of indices can be derived by repeated use of
the BCJ relation and/or the monodromy relation. Therefore, as in the
single color case, we only need to consider the parameter space of one
amplitude, which we can choose to be the symmetric one $\mathcal{A}\left(\left.1234\right|1234\right)$. This has the same form as  (\ref{eq:as})
\begin{equation}
\mathcal{A}\left(\left.1234\right|1234\right)=S\left(\left\{ k_{i},\epsilon_{i}\right\} \right)A\left(s,t\right)\, .
\end{equation}
First of all, under permutations of the external legs, we have
\begin{align}
S\left(\left\{ k_{i},\epsilon_{i}\right\} \right)A\left(s,t\right) & =\mathcal{A}\left(\left.1234\right|1234\right)\,,\\
S\left(\left\{ k_{i},\epsilon_{i}\right\} \right)A\left(u,t\right) & =\mathcal{A}\left(\left.1324\right|1324\right)\,,\\
S\left(\left\{ k_{i},\epsilon_{i}\right\} \right)A\left(s,u\right) & =\mathcal{A}\left(\left.2134\right|2134\right)\,,
\end{align}
by unifying the first color order by applying BCJ relation:
\begin{align}
\frac{u}{t}\mathcal{A}\left(\left.2134\right|2134\right) & =\mathcal{A}\left(\left.1234\right|2134\right)\,,\\
\frac{u}{s}\mathcal{A}\left(\left.1324\right|1324\right) & =\mathcal{A}\left(\left.1234\right|1324\right)\,,
\end{align}
and comparing with (\ref{eq:monodromy}), we arrive at
\begin{equation}
\frac{u}{s}A\left(s,u\right)+e^{i \pi s}A\left(s,t\right)+e^{-i\pi u}\frac{t}{s}A\left(u,t\right)=0\label{eq:bms}\,.
\end{equation}
The Laurent expansion of $A\left(s,t\right)$ is
\begin{equation}
A\left(s,t\right)=-\left(\frac{1}{s}+\frac{1}{t}\right)+\sum_{k\geq q\geq0}g_{k,q}s^{k-q}t^{q}\label{eq:ALE}\,,
\end{equation}
and by solving the monodromy relation we find the coefficients:
\begin{align}
 & \left(\begin{array}{cccccc}
g_{00}\\
g_{1,0} & g_{1,1}\\
g_{2,0} & g_{2,1} & g_{2,2}\\
g_{3,0} & g_{3,1} & g_{3,2} & g_{3,3}\\
g_{4,0} & g_{4,1} & g_{4,2} & g_{4.3} & g_{4,4}
\end{array}\right)\nonumber\\
= & \left(\begin{array}{cccccc}
0\\
\zeta\left(2\right) & \zeta\left(2\right)\\
g_{2,0} & 2g_{2,0} & g_{2,0}\\
\zeta\left(4\right) & \frac{5}{4}\zeta\left(4\right) & \frac{5}{4}\zeta\left(4\right) & \zeta\left(4\right)\\
g_{4,0} & -\zeta\left(2\right)g_{2,0}+3g_{4,0} & -2\zeta\left(2\right)g_{2,0}+4g_{4,0} & -\zeta\left(2\right)g_{2,0}+3g_{4,0} & g_{4,0}\,.
\end{array}\right)\label{eq:bcc}
\end{align}
Note that in (\ref{eq:bms}) we are in fact solving for the Laurent expansion of $\left(s+t\right)A\left(s,t\right)$ for $A\left(s,t\right)$ in (\ref{eq:ms}), so the parameters in (\ref{eq:bcc}) and (\ref{eq:agkq}) are in one-to-one correspondence as expected. Naturally the correspondence between monomials of odd zeta values and free parameters emerges in the bicolor case as well.

\section{Intersection of monodromy plane and the EFThedron}

In this section, we investigate the allowed  space of Laurent coefficients of four-point amplitudes under the combined constraints following from the monodromy relation (\ref{eq:ms}) and unitarity in section \ref{sEFT}. First we demonstrate that monodromy and positivity of Hankel matrices rule out the $t/s$ and $s/t$ poles that correspond to vector exchange. Next, we apply the full EFThedron constraints on the remaining monodromy plane of the single color, bicolor, and finally closed string EFT amplitudes. 
\subsection{The absence of isolated massless poles}
In Section \ref{monosection} it was stated that monodromy relations, when combined with unitarity, enforce the massless pole structure of $A\left(s,t\right)$ to be only of the form $\frac{1}{st}$. Here we give the derivation. Starting with the following Laurent expansion of the four-point amplitude:
\eq
A\left(s,t\right)=-\frac{1}{st}+c\left(\frac{t}{s}+\frac{s}{t}\right)+\frac{b}{s}+\frac{b}{t}+\sum_{k\geq q\geq0}g_{k,q}s^{k-q}t^{q}\,.
\eqe
We first impose the monodromy relation to the above, leading to the following solutions for the couplings up to $k=4$:
\begin{align}
k={-}1:&\quad b = 0\nonumber\\
k= 0:&\quad g_{0,0} = \zeta\left(2\right)+c\nonumber\\
k=2:&\quad g_{2,0} = g_{2,2} =\zeta\left(4\right)-\zeta\left(2\right)c\nonumber\\
&\quad g_{2,1} =\frac{1}{4}\zeta\left(4\right)-\zeta\left(2\right)c\nonumber\\
k=3:&\quad g_{3,1} = g_{3,2} =2g_{3,0}-\zeta\left(2\right)g_{1,0}\nonumber\\
k=4:&\quad g_{4,0}=g_{4,4}=\zeta\left(6\right)-\zeta\left(4\right)c\nonumber\\
&\quad g_{4,2}=-\frac{1}{16}\zeta\left(6\right)+\frac{1}{4}\zeta\left(4\right)c+2g_{4,1}\,. \label{eq:monog} 
\end{align}
Once again one sees that $b$ is set to zero by monodromy relations alone. Furthermore, $g_{2n,0}$ is solved in terms of $c$ alone and has the following general form:
\eq
g_{2n,0}=\zeta\left(2n+2\right)-\zeta\left(2n\right)c\,.
\eqe
We consider the minor of Hankel matrix (\ref{hmc}):
\eq
H_{1,N\times N}=\left(\begin{array}{cccc} g_{0,0} & g_{2,0} & \cdots & g_{2N-2,0} \\ g_{2,0} & g_{4,0} & \cdots & g_{2N,0} \\ \vdots & \vdots & \ddots & \vdots \\ g_{2N{-}2,0} & g_{2N,0} & \cdots & g_{4N{-}4,0} \end{array}\right)\,,\quad 
H_{2,N\times N}=\left(\begin{array}{cccc} g_{2,0} & g_{4,0} & \cdots & g_{2N,0} \\ g_{4,0} & g_{6,0} & \cdots & g_{2N{+}2,0} \\ \vdots & \vdots & \ddots & \vdots \\ g_{2N,0} & g_{2N+2,0} & \cdots & g_{4N{-}2,0} \end{array}\right)\,.
\eqe
The positive condition $\textrm{det}(H_i)>0$ up to $N=30$ implies the condition ${-}4.24\times 10^{-6}<c<6.81\times 10^{-6}$, so we can conclude $c$ is asymptotically fixed to zero. 

Thus we conclude that we can simply begin with 
\eq
A\left(s,t\right)=-\frac{1}{st}+\sum_{k\geq q\geq0}g_{k,q}s^{k-q}t^{q}\,,
\eqe
and study the intersection geometry further.

\subsection{Combined constraints for single color order amplitude}

In Section \ref{sEFT} we showed that unitarity implies two different types of constraints: being inside the cyclic polytope (\ref{cpc}), and the positivity of Hankel matrices (\ref{hmc}),(\ref{phc}), which we already used in ruling out the $\frac{s}{t}$ and $\frac{t}{s}$ terms. In the parameter space for all independent $g_{k,q}$, these unitary constraints carve out the EFThedron, a positive region bounded by a set of codimension one surfaces. On the other hand, the monodromy constraint fixes a subset of $g_{k,q}$ and imposes linear relations among the unfixed ones, therefore defines a lower dimensional plane, which we call the monodromy plane. The allowed Laurent coefficients must lie in the intersection of  the EFThedron and the monodromy plane. The final shape of the allowed region can be straightforwardly derived by imposing  (\ref{cpc}), (\ref{hmc}) and (\ref{phc}) on this hyperplane.

By setting $c=0$ in (\ref{eq:monog}), we arrive at the following defining relations for the monodromy plane:
\begin{align}
k=0:&\quad\ g_{0,0} = \zeta\left(2\right)\nonumber\\
k=2:&\quad\ g_{2,0} = g_{2,2} = \zeta\left(4\right)\nonumber\\
&\quad\ g_{2,1} = \zeta\left(4\right)/4\nonumber\\
k=3:&\quad\ g_{3,1} = g_{3,2} =2g_{3,0}-\zeta\left(2\right)g_{1,0}\nonumber\\
k=4:&\quad g_{4,0}=g_{4,4}=\zeta\left(6\right)\nonumber\\
&\quad g_{4,2}=-\zeta\left(6\right)/16+2g_{4,1}\,.\label{eq:mono3} 
\end{align} 
where we choose $g_{1,0}$ and $g_{3,0}$ as the free parameters for the hyperplane with $k\leq 3$, and $g_{1,0}$, $g_{3,0}$, and $g_{4,1}$ for $k\leq 4$. In the following we will focus on the two-dimensional reduced space for $(g_{1,0},g_{3,0})$ and the three-dimensional reduced space for $(g_{1,0},g_{3,0},g_{4,1})$. We will derive and illustrate graphically the region carved out by increasing order of unitarity constrains and show how the results leads to our main conjecture.
\subsubsection{The $k=3$ Geometry}
We will start by considering the unitarity constrains only up to the order $k=3$. First we consider the the cyclic polytope constraint for $k=2$, which will bound $\vec{g}_2=(1,g_{2,1}/g_{2,0},g_{2,2}/g_{2,0})$. From cyclic symmetry we have $g_{2,1}=g_{2,2}$, and so $\vec{g}_2$  is one parameter vector with 3 components. We have the strongest condition from: $\langle \vec{g}_2\ \vec{\mathcal{G}}_1\ \vec{\mathcal{G}}_2 \rangle >0$. The vector $\vec{\mathcal{G}}_\ell$ of the Gegenbauer polynomial  $G_{\ell}^n(\textrm{cos}\ \theta)$ reads, for dimensions 4, 10 and 26:
\begin{align}
d=4&: \vec{\mathcal{G}}_1=(1,2,0),\quad \vec{\mathcal{G}}_2=(1,6,6)\,,\nonumber\\\nonumber\\
d=10&:\vec{\mathcal{G}}_1=(7,14,0),\quad \vec{\mathcal{G}}_2=(28,126,126)\,,\nonumber\\\nonumber\\
d=26&: \vec{\mathcal{G}}_1=(23,46,0),\quad \vec{\mathcal{G}}_2=(276,1150,1150)\, .
\end{align}
For $k{\,=\,}3$, $\vec{g}_3=(1,g_{3,1}/g_{3,0},g_{3,2}/g_{3,0},g_{3,3}/g_{3,0})$ is still a one parameter vector, with 4 components. The strongest condition is given by considering the cyclic polytope in one lower dimension, which means we only keep components $g_{3,q}$ from $q{\,=\,}0$ to $2$. We use prime to denote this vector $\vec{g}'_3=(1,g_{3,1}/g_{3,0},g_{3,2}/g_{3,0})$ and the extremal condition is: $\langle \vec{g}'_3\ \vec{\mathcal{G}}_1\ \vec{\mathcal{G}}_2 \rangle >0$. 
As a result, the two cyclic polytope constraints for dimensions $4,10,26$ are  as follows\footnote{Note that the cyclic polytope constraints follow from the positivity of coefficients for each Gegenbauer polynomial,  therefore are dimension dependent.}:
 \begin{align}
d=4:& \ 0 \leq \frac{g_{2,1}}{g_{2,0}} \leq \frac{8}{3},\quad 0 \leq \frac{g_{3,1}}{g_{3,0}} \leq 6\,,\nonumber\\\nonumber\\
d=10:& \ 0 \leq \frac{g_{2,1}}{g_{2,0}} \leq \frac{23}{9}, \quad 0 \leq \frac{g_{3,1}}{g_{3,0}} \leq \frac{9}{2}\,,\nonumber\\\nonumber\\
d=26:& \ 0 \leq \frac{g_{2,1}}{g_{2,0}} \leq \frac{63}{25},\quad 0 \leq \frac{g_{3,1}}{g_{3,0}} \leq \frac{25}{6}\label{eq:cyclic3}\, .
\end{align}

The (independent) Hankel matrix constraints up to order $k=3$ consist of the positivity of all $g_{k,q}$, as well as positivity of the determinants of the following three matrices:
\begin{equation}
H_1 =
\begin{pmatrix}
g_{0,0} & g_{1,0}\\
g_{1,0} & g_{2,0}
\end{pmatrix},\ 
H_2 =
\begin{pmatrix}
g_{1,0} & g_{2,0}\\
g_{2,0} & g_{3,0}
\end{pmatrix},\ 
H_3 =
\begin{pmatrix}
g_{1,1} & g_{2,1}\\
g_{2,1} & g_{3,1}
\end{pmatrix}\ ,
\end{equation}
which reduce to the following constraints on the Laurent coefficients
\begin{align}
\det\left(H_1\right) =& \ g_{2,0}g_{0,0} -g_{1,0}^2> 0\,,\nonumber \\
\det\left(H_2\right)  = &\ g_{1,0}g_{3,0}-g_{2,0}^2 > 0 \,,\nonumber\\
\det\left(H_3\right) =& \ g_{1,0}g_{3,1}-g_{2,1}^2 > 0\label{eq:Hankel3}\,.
\end{align}

\begin{figure}[H]
  \centering
  \begin{subfigure}[b]{0.4\linewidth}
    \includegraphics[scale=0.4]{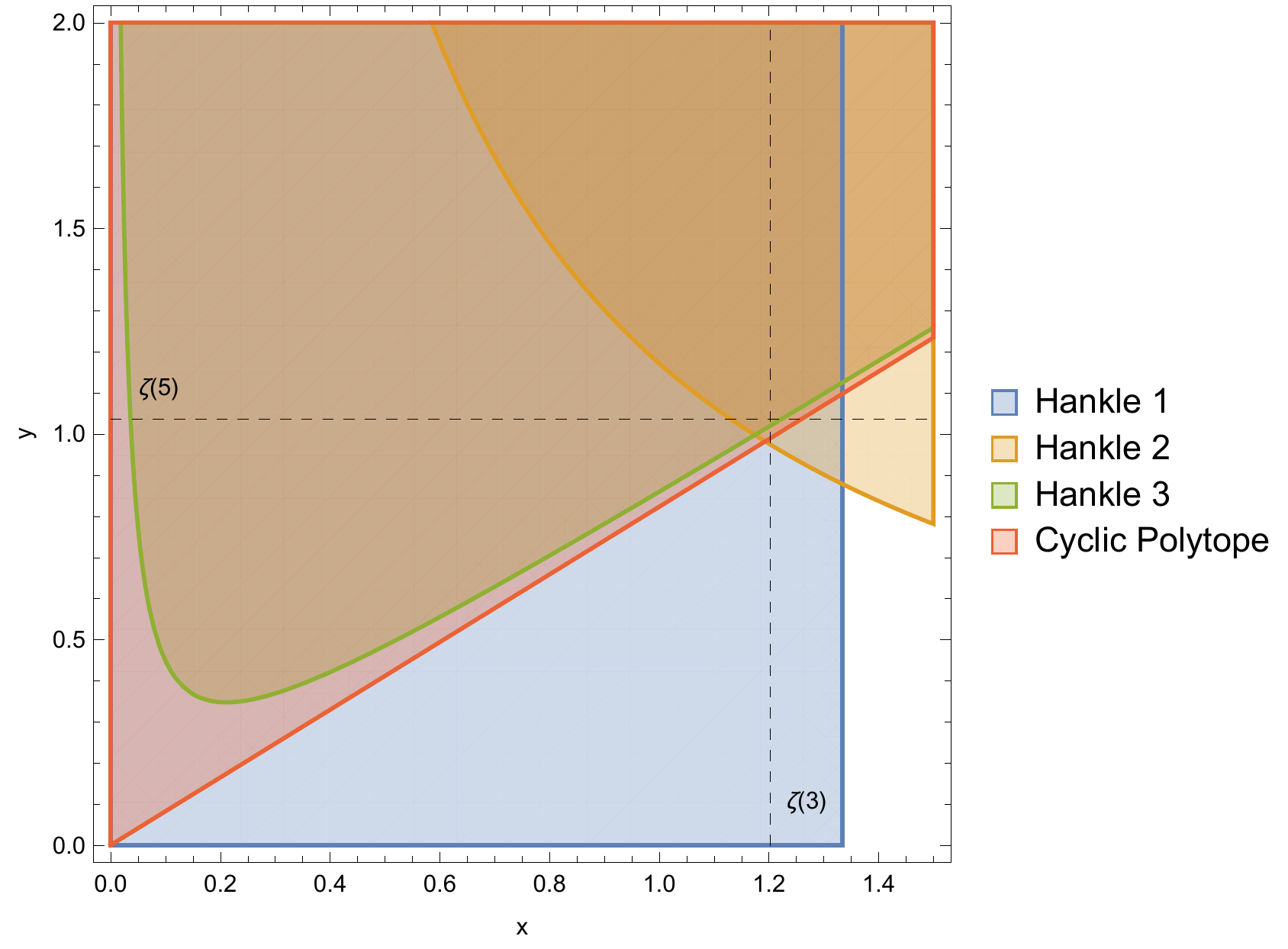}
    \caption{Hankel matrix and cyclic polytope constraints up to $k{=}3$}\label{U3}
  \end{subfigure}
  \begin{subfigure}[b]{0.4\linewidth}
    \includegraphics[scale=0.4]{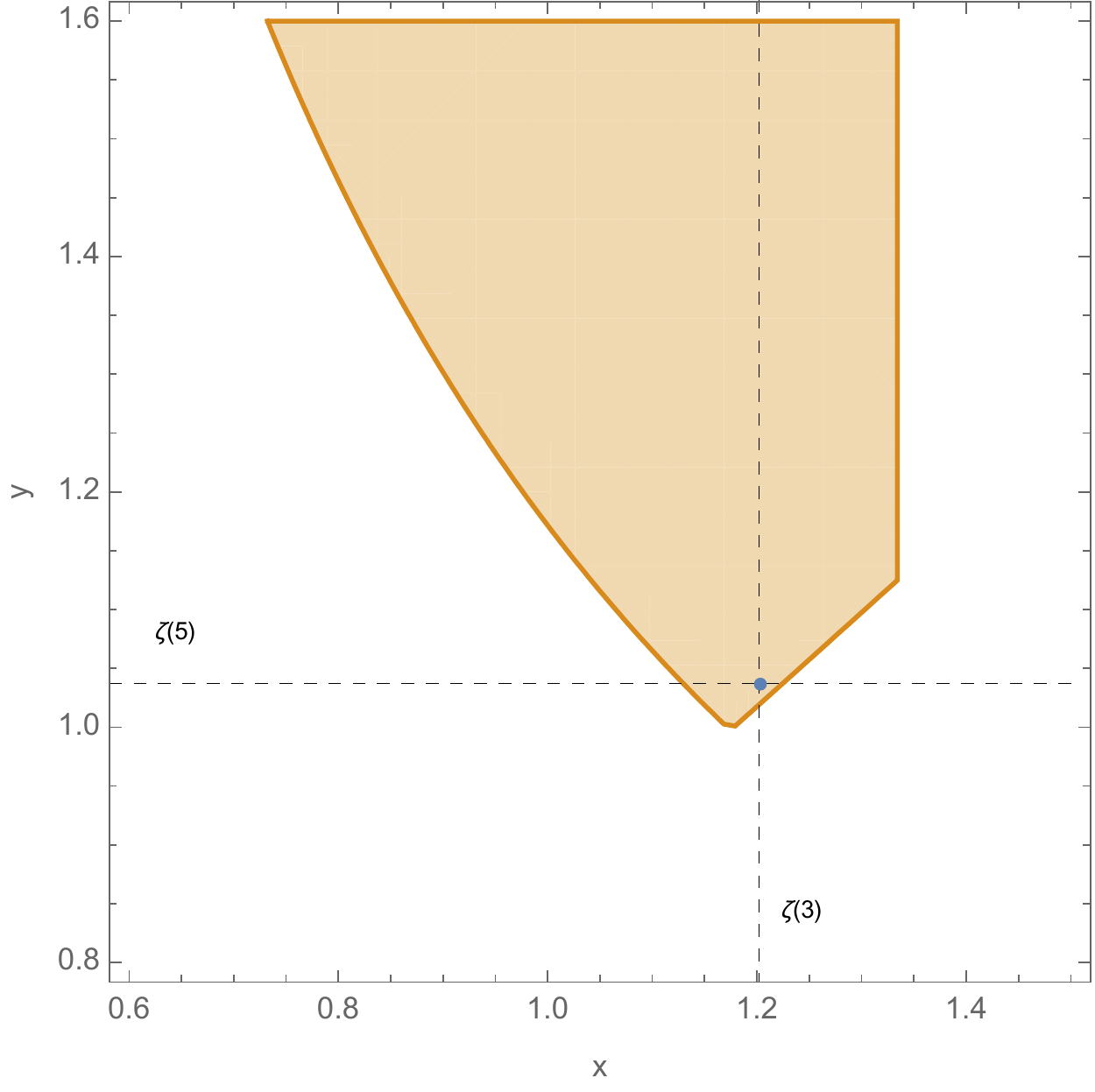}
    \caption{Intersection region and superstring solution }\label{U3c}
    \label{U3c}
  \end{subfigure}
  \caption{Region  of parameter space allowed by monodromy and unitarity up to $k{=}3$
   }
\end{figure}

In the $g_{k,q}$ space of parameters, the monodromy plane is defined by the linear relations in (\ref{eq:mono3}).  In the subspace of  $g_{k,q}$ for $k\leq3$, the monodromy plane is a two-dimensional plane parameterized by coordinates $(x=g_{1,0},y=g_{3,0})$, with explicit coordinate representation:
 \begin{equation}
\left(g_{0,0},g_{1,0},g_{2,0},g_{2,1},g_{3,0},g_{3,1}\right)=\left(\zeta(2),x,\zeta(4),\zeta(4)/4,y,2y-\zeta(2)x\right).\label{2Dmp}
\end{equation}
On this plane, positivity of $g_{1,0}$ and $g_{3,0}$ translates to positivity of coordinates, and the space-time dimensional independent part of cyclic polytope constraints (\ref{eq:cyclic3}) reduces to a linear inequality 
\begin{align}
y>\frac{\zeta(2)}{2}x\,,
\end{align}
 while the dimensional dependent part is trivial, for positive $x$ and $y$. Finally, the positive Hankel matrix determinants in (\ref{eq:Hankel3}) reduce to the following quadratic constraints on $x$ and $y$:
\begin{align}
\det\left(H_1\right) =& -x^2+\zeta(4)\zeta(2) > 0\,,\nonumber \\
\det\left(H_2\right)  = &\ xy-\zeta(4)^2 > 0\,, \nonumber\\
\det\left(H_3\right) =& -\zeta(2)x^2+2xy-\frac{1}{16} \zeta(4)^2 > 0\label{eq:RHankel3}\,.
\end{align}
The constraints are demonstrated graphically in Figure \ref{U3}, and  their intersection is  magnified in Figure \ref{U3c}, along with a marked point corresponding to the open string solution of $(x,y)$.

\subsubsection{The $k=4$ Geometry}
Moving on to $k=4$ geometry, we demonstrate the crucial feature that at fixed $k$, including Hankel matrices (\ref{hmc}) constraints of higher order $k$ not only bounds the new degrees of freedom, it further constrains the low order parameters. For instance, as monodromy requires $ g_{4,0}=\zeta\left(6\right)$, we can improve the Hankel matrix constraints   for $\left\{g_{k,q}|k\leq 3 \right\}$ by including the Hankel matrices involving $\left\{g_{4,0}\right\}$
\begin{equation}
H_4 =
\begin{pmatrix}
g_{0,0} & g_{1,0} & g_{2,0}\\
g_{1,0} & g_{2,0} & g_{3,0}\\
g_{2,0} & g_{3,0} & g_{4,0}
\end{pmatrix},\,
H_5 =
\begin{pmatrix}
g_{2,0} & g_{3,0}\\
g_{3,0} & g_{4,0}
\end{pmatrix},\,
H_6 =
\begin{pmatrix}
g_{0,0} & g_{2,0}\\
g_{2,0} & g_{4,0}
\end{pmatrix}\,,
\end{equation}
which leads to two new constraints on the monodromy plane (\ref{2Dmp}) :
\begin{align}
\det\left(H_4\right) = -\frac{4}{7}\zeta(4)x^2+\frac{4}{5}\zeta(2)xy-y^2+\frac{3}{25}\zeta(2)^2\zeta(6) > 0 \,,\label{eq:Hankel4}
\end{align}
and
\begin{align}
\det\left(H_5\right) = \zeta (4)\zeta (6)-y^2 > 0 \label{eq:Hankel5}\,.
\end{align}

\begin{figure}[H]
  \centering
  \begin{subfigure}[b]{0.41\linewidth}
    \includegraphics[scale=0.41]{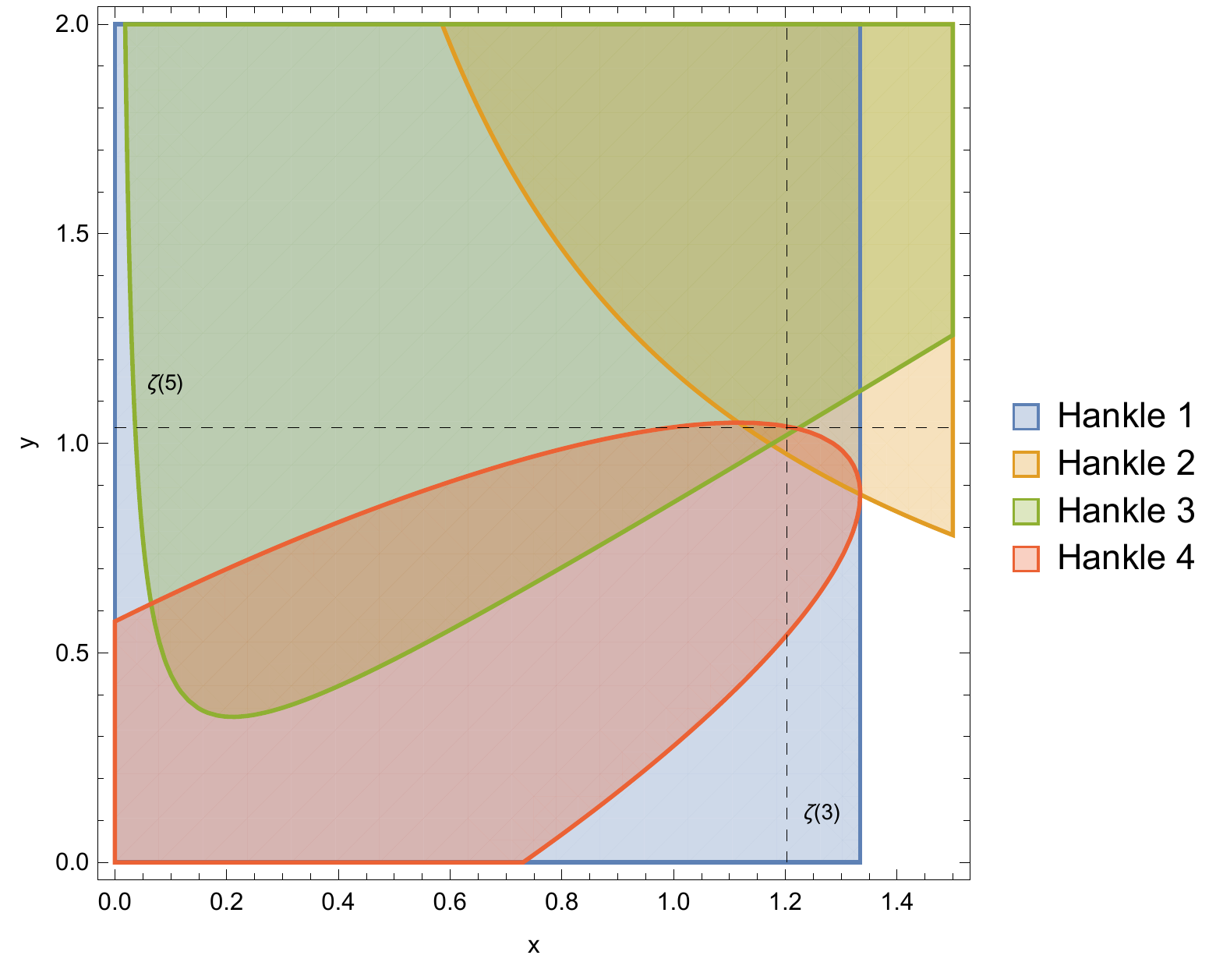}
    \caption{Remaining constraints after reduction up to $k{=}4$}\label{subfig:hankel4}
  \end{subfigure}
  \begin{subfigure}[b]{0.41\linewidth}
    \includegraphics[scale=0.41]{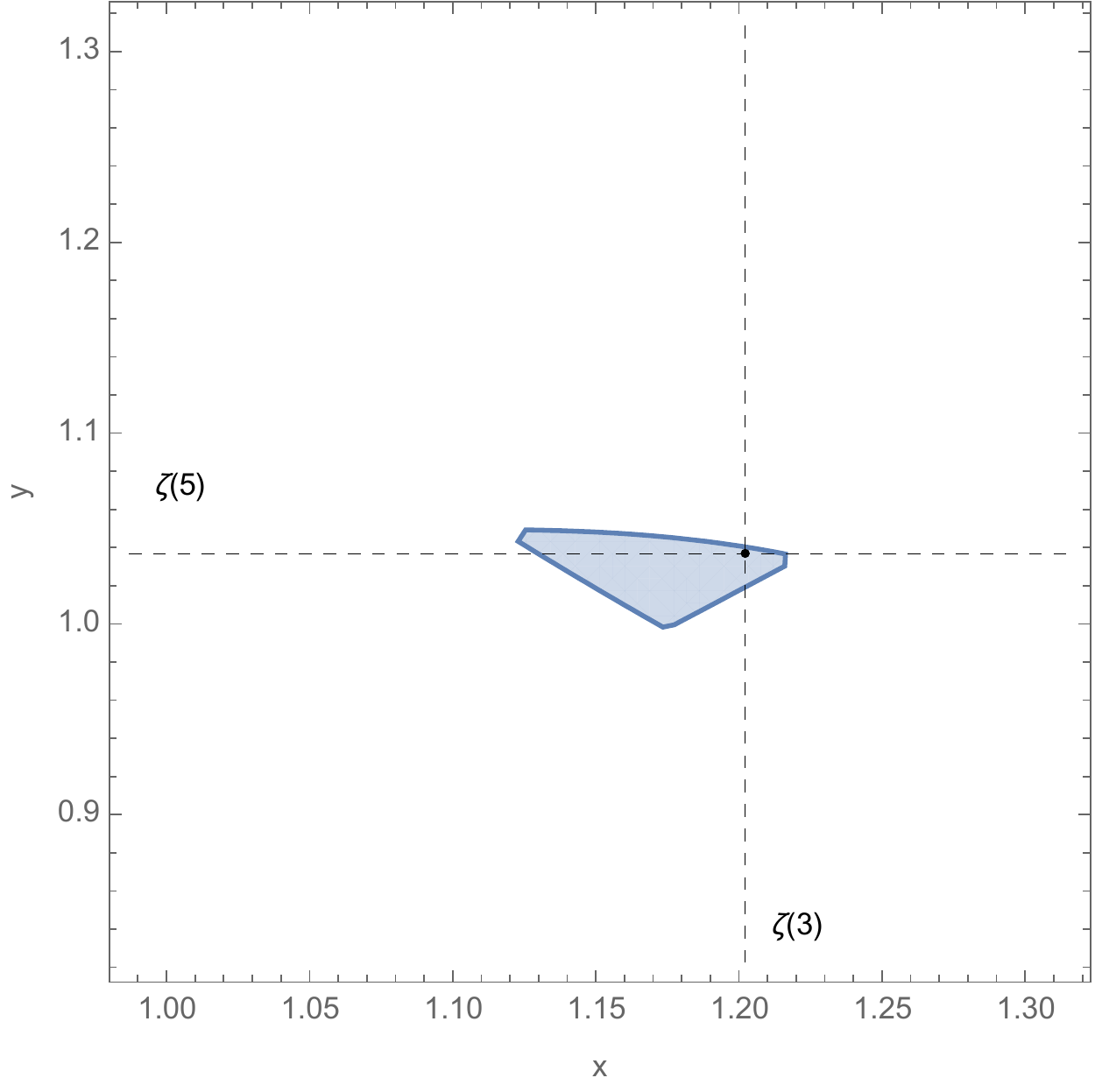}
    \caption{Intersection region and superstring solution }
    \label{U4s}\label{subfig:intersection2d}
  \end{subfigure}
 \begin{subfigure}[b]{0.41\linewidth}
  \includegraphics[scale=0.45]{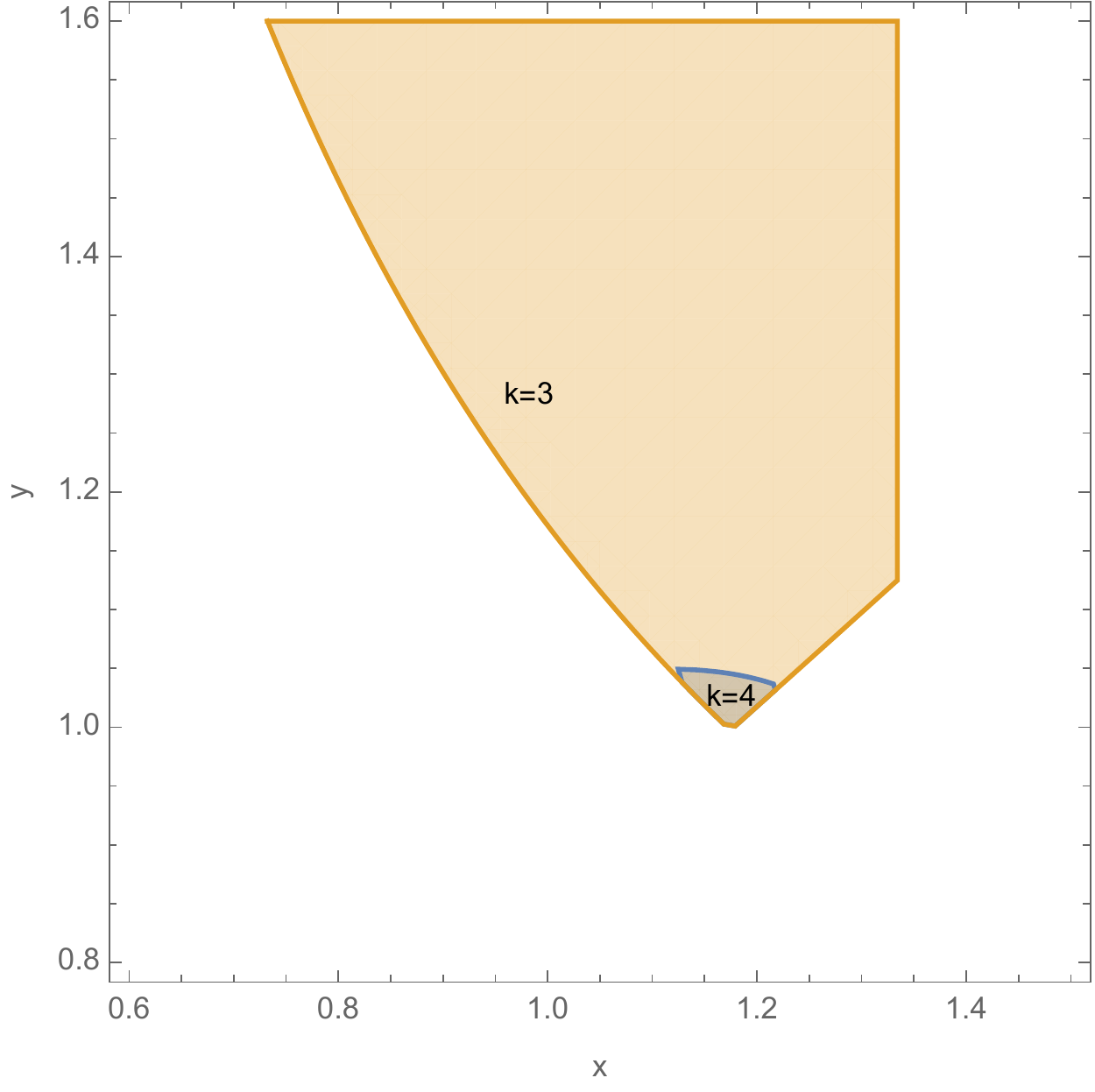}
\caption{$k=3$ and $k=4$ constraints}
\label{Uc34}
  \end{subfigure}
  \caption{Region  of parameter space allowed by monodromy and unitarity up to $k{=}4$ 
   }
\end{figure}

Next we consider the polytope constraint ($\ref{cpc}$) of the vector
\eq
\vec{g}_4=(1,g_{4,1}/g_{4,0},g_{4,2}/g_{4,0},g_{4,3}/g_{4,0},g_{4,4}/g_{4,0})\, ,
\eqe
which has two independent variables due to the symmetry  $g_{k,q}=g_{k,k-q}$,
\eq
\langle \vec{g}_4\ \vec{\mathcal{G}}_0\ \vec{\mathcal{G}}_i\ \vec{\mathcal{G}}_{i+1}\ \vec{\mathcal{G}}_j\ \vec{\mathcal{G}}_{j+1} \rangle >0\label{C4}\,,
\eqe
and positive determinants for the following matrices
\begin{equation}
H_7 =
\begin{pmatrix}
g_{2,1} & g_{3,1}\\
g_{3,1} & g_{4,1}
\end{pmatrix},\ 
H_8 =
\begin{pmatrix}
g_{2,2} & g_{3,2}\\
g_{3,2} & g_{4,2}
\end{pmatrix}\,.
\end{equation}
The monodromy plane in the space of $g_{k,q}$ for $k{\,\leq\,}4$ is three dimensional and parameterized by $(x,y,z)=(g_{1,0},g_{3,0},g_{4,1})$, where positive determinants of $H_7$, $H_8$ lead to the following constraints
\begin{align}
\textrm{det}(H_7)>0\quad:&\quad\frac{\zeta\left(4\right)}{4}z-\left(2y-\zeta\left(2\right)x\right)^{2}>0\,,\nonumber\\
\textrm{det}(H_8)>0\quad:&\quad\zeta\left(4\right)\left(2z-\frac{\zeta\left(6\right)}{16}\right)-\left(2y-\zeta\left(2\right)x\right)^{2}>0\label{H7H8}\,.
\end{align} 
At this level, the product Hankle matrices  introduced in ($\ref{phc}$) also lead to non-trivial constraints.
For space-time dimension $d=4$, two such product Hankle matrices are given by:
\eq
H_{p,1}=
\begin{pmatrix}
\langle\vec{g}_{2}\ \vec{\mathcal{G}}_{2}\ \vec{\mathcal{G}}_{3}\rangle & \langle\vec{g}_{3}\ \vec{\mathcal{G}}_{2}\ \vec{\mathcal{G}}_{3}\rangle\\
\langle\vec{g}_{3}\ \vec{\mathcal{G}}_{2}\ \vec{\mathcal{G}}_{3}\rangle & \langle\vec{g}_{4}\ \vec{\mathcal{G}}_{2}\ \vec{\mathcal{G}}_{3}\rangle
\end{pmatrix},\quad
H_{p,2}=
\begin{pmatrix}
\langle\vec{g}_{2}\ \vec{\mathcal{G}}_{3}\ \vec{\mathcal{G}}_{4}\rangle & \langle\vec{g}_{3}\ \vec{\mathcal{G}}_{3}\ \vec{\mathcal{G}}_{4}\rangle\\
\langle\vec{g}_{3}\ \vec{\mathcal{G}}_{3}\ \vec{\mathcal{G}}_{4}\rangle & \langle\vec{g}_{4}\ \vec{\mathcal{G}}_{3}\ \vec{\mathcal{G}}_{4}\rangle
\end{pmatrix}\,,
\eqe
implying the following conditions: 
\begin{align}
\textrm{det}(H_{p,1})=&\ 41\pi^6-75\pi^4 x^2-43200 y^2-720\pi^2(5xy+z)>0\,,\nonumber\\
 \textrm{det}(H_{p,2})=&\ 64801\pi^6-60(845\pi^4x^2+1590480y^2+3\pi^2(24440xy+5203z))>0\label{Hp1Hp2}\,.
\end{align}
Despite the fact that (\ref{H7H8},\ref{Hp1Hp2}) are constraints for parameters $(x,y,z)$, the region they carve out can be projected to the two-dimensional plane of $(x,y)$. The $k\leq 4$ constraints (\ref{eq:cyclic3},\ref{eq:RHankel3},\ref{eq:Hankel4},\ref{eq:Hankel5}) and the projection of (\ref{C4},\ref{H7H8},\ref{Hp1Hp2}) have significant overlap, so the total allowed region straightforwardly reduces to the intersection of the previous region (\ref{eq:RHankel3}) and (\ref{eq:Hankel4}), shown in Figure \ref{subfig:hankel4} and magnified in Figure \ref{subfig:intersection2d}. Evidently, as shown in Figure \ref{Uc34}, the $k=4$ allowed region for $x,y$ is notably smaller  than the $k=3$ one.

 At this stage, $k=4$, we are able to  fix the coordinates $x$ and $y$ on the monodromy plane to within 10\%  accuracy around the actual values of $\zeta(3)=1.20\ldots$ and $\zeta(5)=1.03\ldots$ respectively. Moreover, the string value appears to be sitting very close the boundary of the new allowed region.\footnote{Note that while the massless bosonic string amplitude, for example~\cite{Huang:2016tag} 
\eq
\mathcal{A}_{Bosonic}(1^{+},2^{+},3^{+},4^{+})=\frac{\left[12\right]\left[34\right]}{\left\langle 12\right\rangle\left\langle 34\right\rangle}stu\left(1{-}\frac{1}{s{+}1}{-}\frac{1}{t{+}1}{-}\frac{1}{u{+}1}\right)\frac{\Gamma(-s)\Gamma(-t)}{\Gamma(1-s-t)}
\eqe
also satisfies monodromy relations (\ref{eq:ms}), the presence of Tachyon state will automatically result in violation of the Hankel matrix bound.}

\begin{figure}[h!]
  \centering
  \begin{subfigure}[b]{0.45\linewidth}
    \includegraphics[scale=0.4]{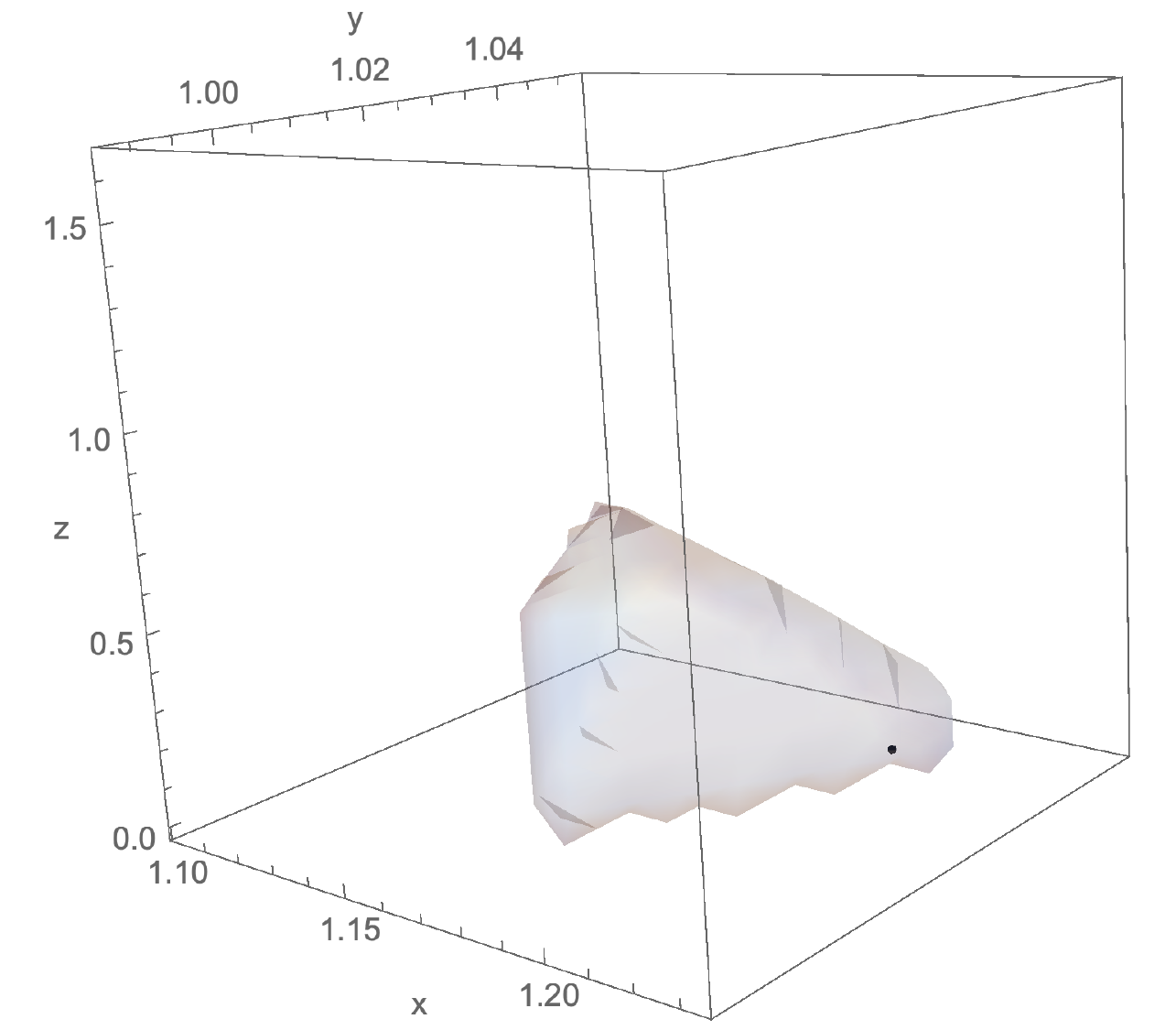}
    \caption{$d=4$}\label{U43Da}
  \end{subfigure}
  \begin{subfigure}[b]{0.45\linewidth}
    \includegraphics[scale=0.4]{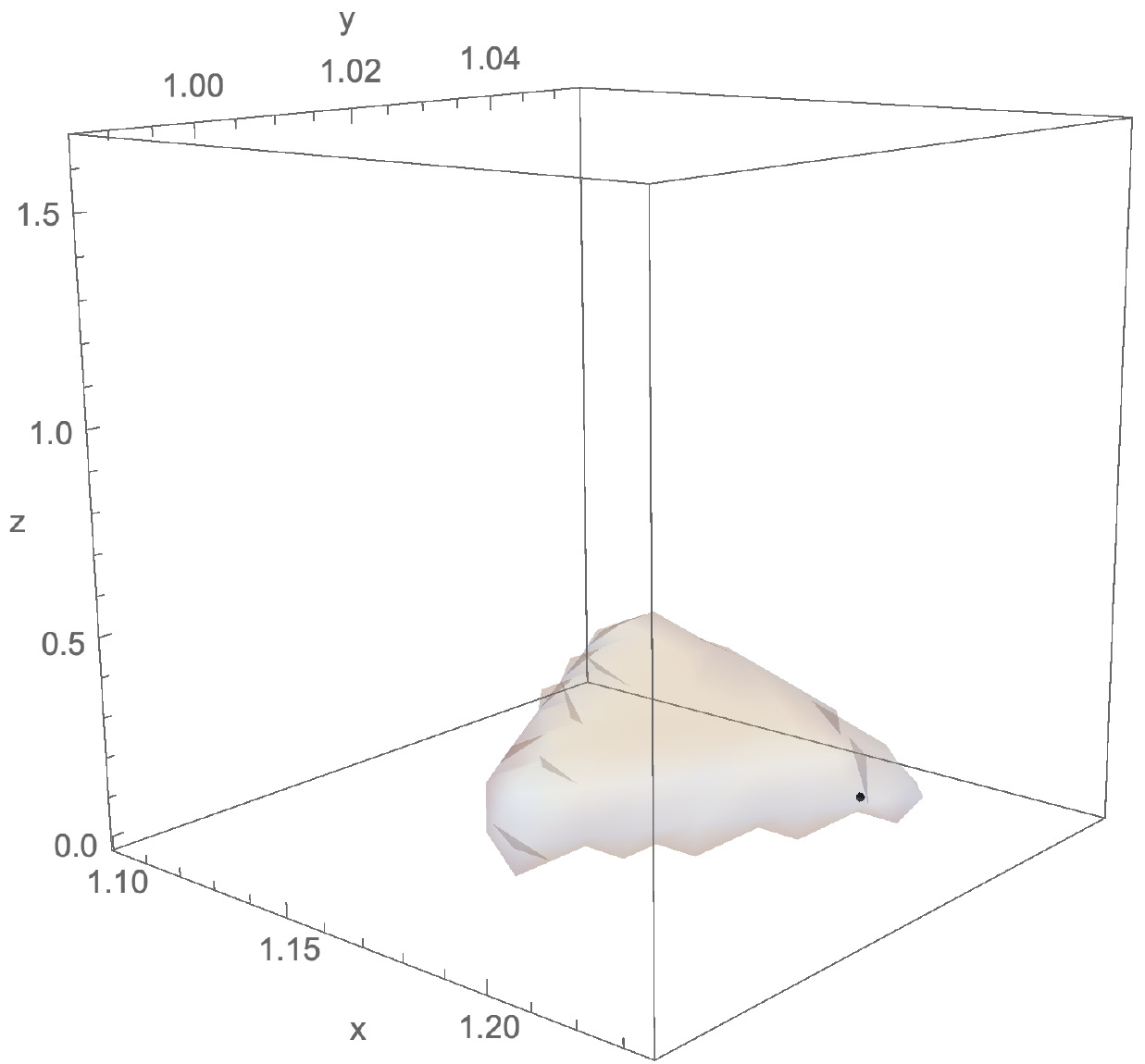}
    \caption{$d=26$}
    \label{U43Db}
  \end{subfigure}

  \caption{Region of parameter space allowed by monodromy and unitarity up to $k{=}4$, and a point corresponding to the open string solution. }
  \label{U43D}
\end{figure}

 We now return to the three-dimensional monodromy plane defined by $(x,y,z)=(g_{1,0},g_{3,0},g_{4,1})$. An illustration for the three-dimensional intersection of all unitary constraints up to order $k=4$,   (\ref{eq:cyclic3},\ref{eq:RHankel3},\ref{eq:Hankel4}) as well as (\ref{eq:Hankel5},\ref{C4},\ref{H7H8},\ref{Hp1Hp2}), is presented in Figure \ref{U43Da} and Figure \ref{U43Db} for space-time dimension $d{=}4$ and $d{=}26$ respectively. It will serve as a starting data set for further investigations of compatibility with higher order unitary constraints.

\subsubsection{Peeking from beyond $k=4$}
Previously, we have seen that by considering $k=4$ constraints we further reduce the allowed region in $k\leq 3$. Here we will do the same at higher orders, to further restrict the region for $k=4$. Analytical reduction soon becomes impractical as  often times the constraints are of higher algebraic order, and almost all new constraints needs to be projected. For example, at order $k=6$, we encounter  the positive determinant condition for the Hankel matrix 
\begin{align}H_9=\left(
\begin{array}{cccc}
 g_{0,0} & g_{1,0} & g_{2,0} & g_{3,0} \\
 g_{1,0} & g_{2,0} & g_{3,0} & g_{4,0} \\
 g_{2,0} & g_{3,0} & g_{4,0} & g_{5,0} \\
 g_{3,0} & g_{4,0} & g_{5,0} & g_{6,0} \\
\end{array}
\right)\,, \end{align}
which is an order $3$ inequality of free parameters $g_{1,0}=x$, $g_{3,0}=y$, and $g_{5,0}$. The projection onto the $\left(x,y\right)$-plane is obtained by solving for values of $x,y$ such that there exists a $g_{5,0}$  for which $\det{H_9}>0$, i.e. points that can be uplifted into the $k=5$ geometry. Therefore, for order $k=6$ and beyond, for practical reasons, we will carry out  the constraints in a numerical fashion and will only include the the most relevant subset.

\begin{figure}[h!]
\begin{center}
\includegraphics[scale=0.35]{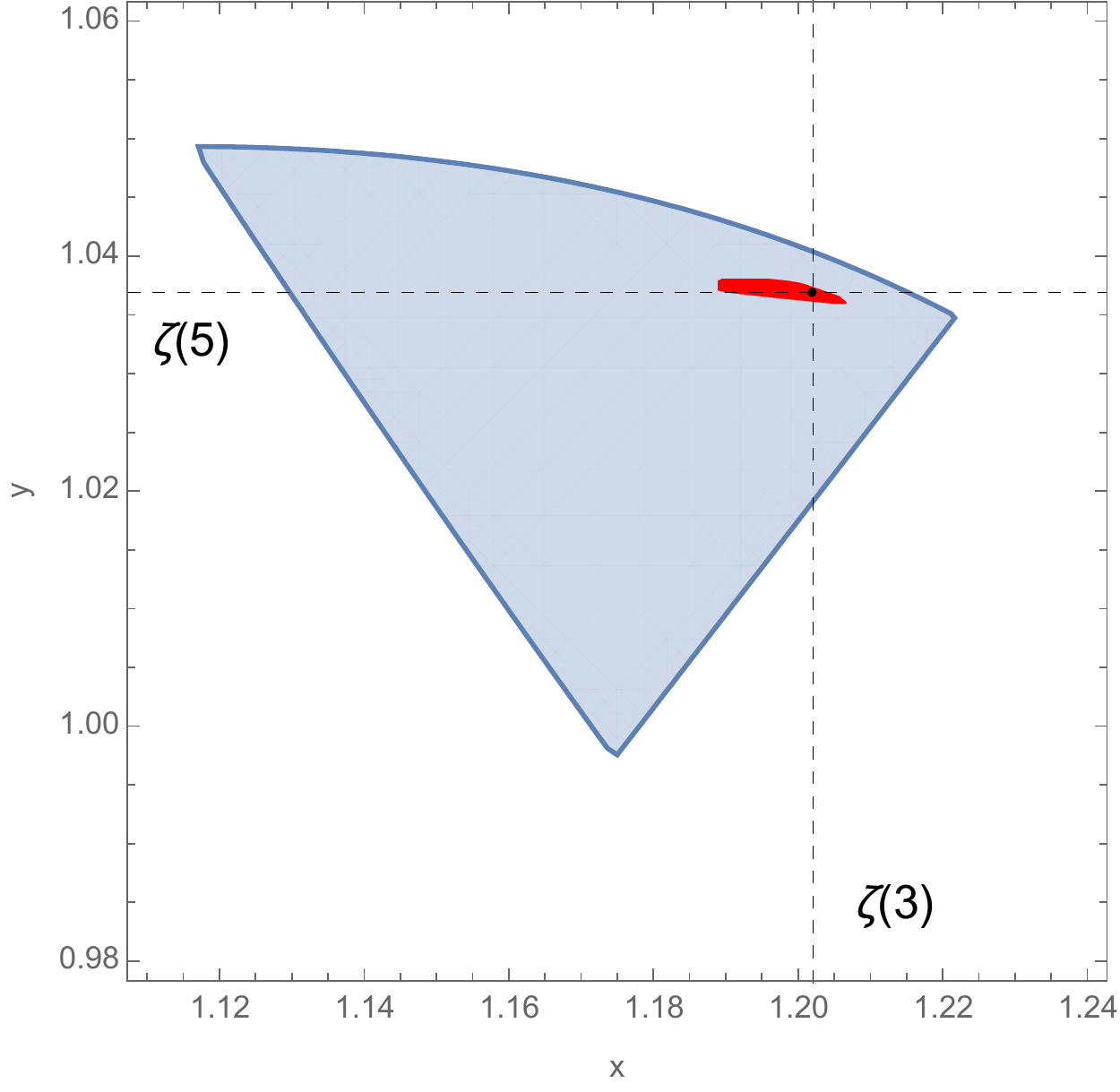}
\caption{Blue region reduces to red region after projecting $\det(H_9)>0$ constraint on $x, y$.}
\label{fig:U4h1}
\end{center}
\end{figure}

Our numerical survey reveals that even the inclusion  of a subset of order $k=6$ constraints can give rise to much stronger conditions compared to the projection of $k=4$ constraints. We impose compatibility with  constraints from the  $k=5$ cyclic polytope,  all Hankel and product Hankel matrices, and one $k=6$ matrix $H_9$ on points in the blue region in Figure \ref{subfig:intersection2d}. 
We are able to rule out most of the region and achieve a much smaller allowed region, marked in red in  Figure \ref{fig:U4h1}. We compare to the original region in blue to manifest the magnitude of this reduction.

\begin{figure}[h!]
  \centering
  \begin{subfigure}[b]{0.4\linewidth}
    \includegraphics[scale=0.45]{U43Dh1.pdf}
  \caption{}\label{fig:U43Dh1}
  \end{subfigure}
  \begin{subfigure}[b]{0.4\linewidth}
    \includegraphics[scale=0.45]{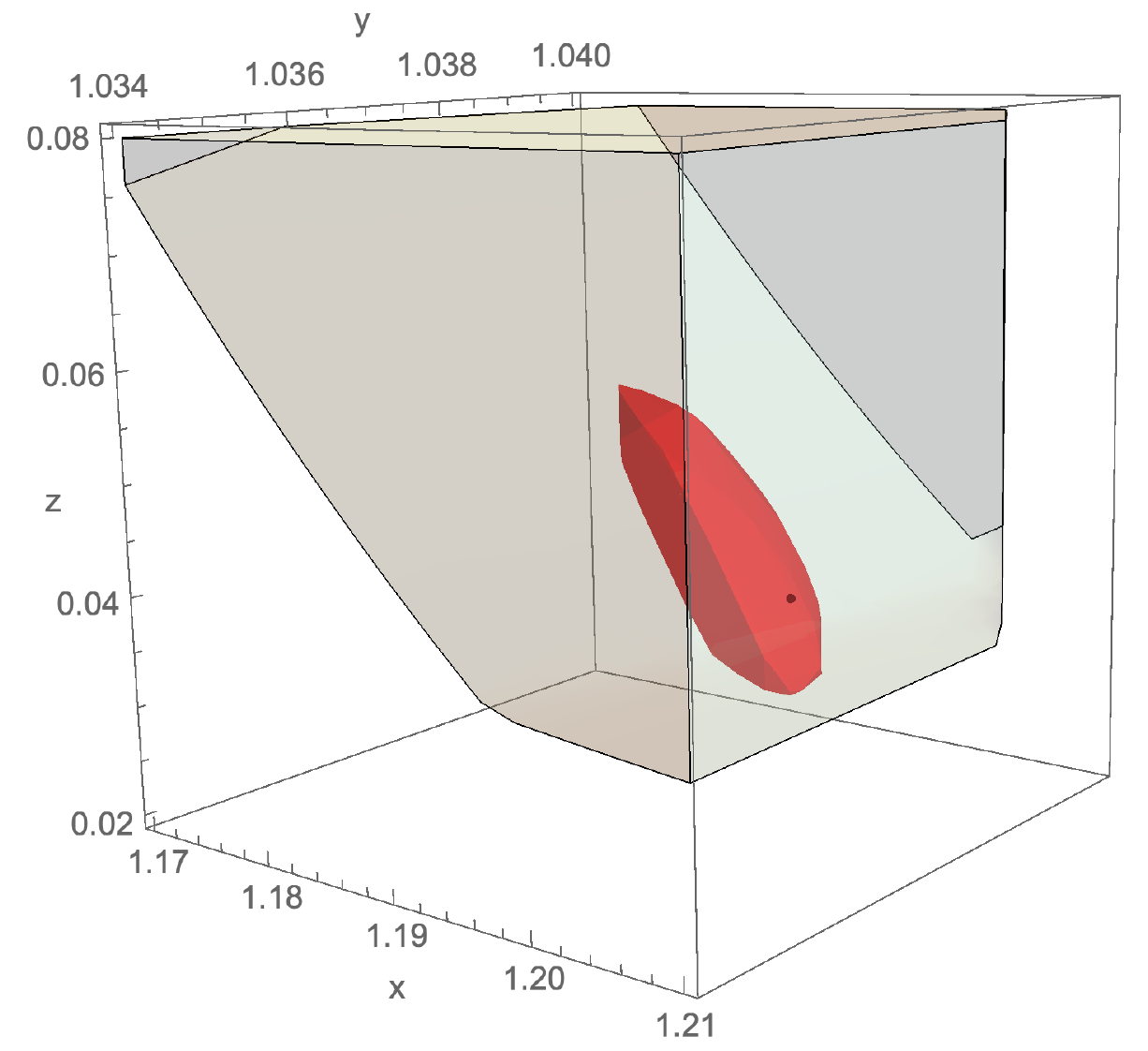}
  \caption{  }\label{fig:U43Dh2}
  \end{subfigure}
  \caption{Region of parameter space fixed by up to $k{=}4$ constraints (gray) and projected $k{=}5$ constraint (red). We can extract the ratio of red region width with respect to actual superstring results in each $(x,y,z)$ direction: $\frac{x^{max}-x^{min}}{x^{string}}\approx1.5\%,\quad \frac{y^{max}-y^{min}}{y^{string}}\approx0.2\%, \quad\frac{z^{max}-z^{min}}{z^{string}}\approx52.8\%$
  }
\end{figure}

We can perform the same scanning for points in Figure \ref{U43D} by imposing compatibility with  $\det(H_9)>0$, as well  as all positivity constrains for $k=5$. The result is again a more constrained region, which we show in red in Figure \ref{fig:U43Dh1}. 
We observe that, by including all constraints up to $k=5$  and one at $k=6$, we have already fixed the value for $(x,y,z)$ to the string value with the following precision:
\eqa
&&\nonumber\frac{x^{max}-x^{min}}{x^{string}}=\frac{1.20667-1.18890}{\zeta(3)}\approx1.5\%,\\
&& \frac{y^{max}-y^{min}}{y^{string}}=\frac{1.03808-1.03594}{\zeta(5)}\approx0.2\%,\nonumber\\
 &&\frac{z^{max}-z^{min}}{z^{string}}=\frac{0.05699-0.03560}{(\pi^6-630\zeta(3)^2)/1260)}\approx52.8\%\,.
\eqae

\begin{figure}[h!]
  \centering
  \begin{subfigure}[b]{0.4\linewidth}
    \includegraphics[scale=0.45]{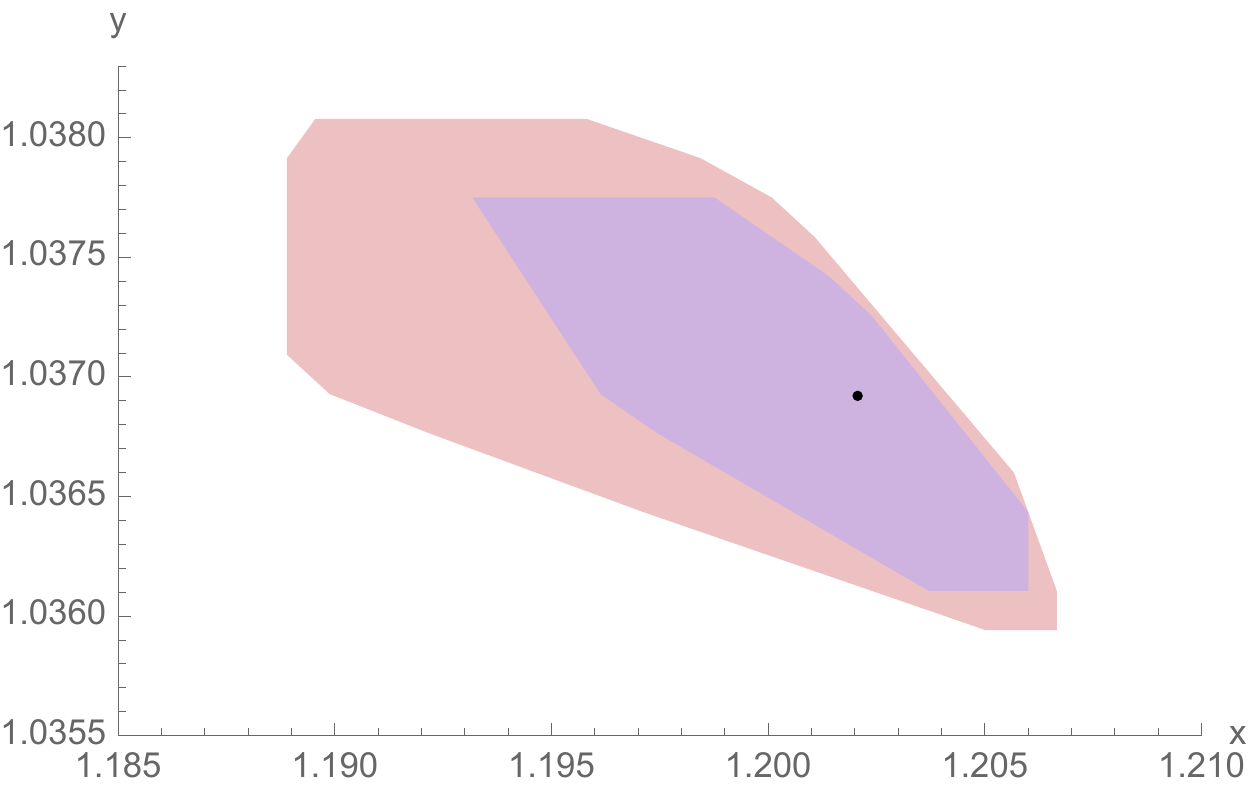}
\caption{Projecting $k=7$ and one $k=8$ constraint on $(x,y)$.}
\label{fig:U3h1}
  \end{subfigure}
  \begin{subfigure}[b]{0.4\linewidth}
    \includegraphics[scale=0.6]{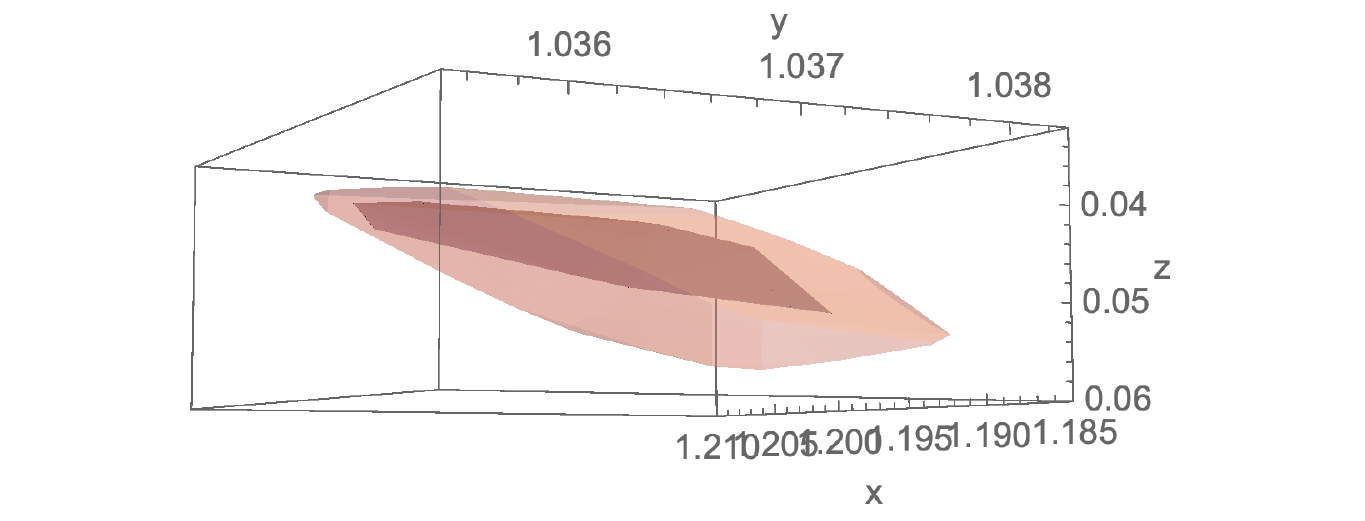}
\caption{Projecting $k=7$ and one $k=8$ constraint on $(x,y,z)$.}\label{Remain}
  \end{subfigure}
  \caption{Region of parameter space fixed by projecting up to $k=7$ and $k=8$ constraints}
  \end{figure}

Extending to $k=8$, the reduction of constraints becomes difficult even with a numerical scan. However, we are able to use the \textbf{FindInstance} function in \emph{Mathematica}  to verify if a point in the $k=6$ allowed region of Figure \ref{fig:U4h1}  or Figure \ref{fig:U43Dh1} is compatible with $k=7,8$ Hankel matrix constraints. We impose the positivity constraints for all $k=7$ Hankel matrices, and the following  $k=8$ principle minor $H_{10}$
 \begin{align}\det(H_{10})=\left(
\begin{array}{ccccc}
 g_{0,0} & g_{1,0} & g_{2,0} & g_{3,0}&  g_{4,0} \\
 g_{1,0} & g_{2,0} & g_{3,0} & g_{4,0} & g_{5,0}\\
 g_{2,0} & g_{3,0} & g_{4,0} & g_{5,0}  & g_{6,0}\\
 g_{3,0} & g_{4,0} & g_{5,0} & g_{6,0}  & g_{7,0}\\
 g_{4,0} & g_{5,0} & g_{6,0}  & g_{7,0} & g_{8,0}\\
\end{array}
\right) >0\,.\end{align}
The allowed region on the $(x,y)$-plane and $(x,y,z)$-plane is now further reduced to the purple region in Figure \ref{fig:U3h1} and Figure \ref{Remain} respectively.

 As we find that the allowed regions of  Laurent coefficients keep shrinking to the string value as we impose increasing order of unitarity constrains, we therefore conjecture that the full four-point open superstring amplitude  $A\left(s,t\right)$ \textit{is uniquely determined by the intersection geometry of the monodromy plane with the EFThedron.}


\subsubsection{Checking a corollary of the main conjecture}
Verifying the conjecture becomes computationally difficult at higher orders in $k$, but we can test a simpler yet non-trivial  corollary, with significant precision. It is a  straightforward consequence of the conjecture that for a EFT amplitude with some of its Laurent coefficients set equal the string value, the remaining unknown parameters can be fixed by monodromy and unitarity to the string value as well.

By setting $(g_{1,0},g_{3,0},g_{4,1})$ to their string values $(\zeta(3),\zeta(5),3/4\zeta(6)-1/2\zeta(3)^2)$, we can search for solutions of the higher order parameters to the $k=7$ EFThedron constraints via \textbf{FindInstance}. We immediately find:
\eq
(g_{5,0},g_{6,1},g_{7,0},g_{7,2})=(1.00834,0.00862,1.00202,0.00035)\,,
\eqe 
matching the string values up to four digits:
\eqa
\nonumber&&\left(\zeta(7),\frac{\pi^8}{7560}{-}\zeta(3)\zeta(5),\zeta(9),\frac{8\pi^6\psi^{(2)}(1){+}9\pi^4\psi^{(4)}(1){+}6\pi^2\psi^{(6)}(1){-}180\psi^{(2)}(1)^3{-}2\psi^{(8)}(1)}{8640}\right)\\
&&=(1.00835,0.00865,1.00201,0.00032)\,.
\eqae

 The most manageable case of this corollary is when all but one parameter is fixed to the string value. For example, at $k{=}5$ monodromy leaves four independent variables. We can deform $g_{1,0}$, and keep the remaining parameters at the superstring value:
\begin{align}
g_{1,0}=&\zeta(3)+x\nonumber\\
g_{3,0}=&\zeta(5)\nonumber\\
g_{4,1}=&\frac{3}{4}\zeta(6)-\frac{1}{2}\zeta(3)^2\nonumber\\
g_{5,0}=&\zeta(7)\,,
\end{align}
including all higher order $k$ couplings. Our conjecture implies that as we increase the order of the constraints, $x$ should be bounded closer and closer to zero. We restrict to just using the following choice of Hankel matrices for simplicity:
\begin{align}
H_{n}=
\begin{pmatrix}
g_{n,i}& g_{n+1,i}\\
g_{n+1,i}& g_{n+2,i}\\
\end{pmatrix}\,,
\end{align}  
for all $i\le n$, and we impose all entries to be positive, along with $\textrm{det} (H_{n})>0$. Even with this drastically reduced set of constraints we find that the value of $x$ quickly becomes highly constrained:
\begin{align}
H_{5}:\ & -10^{-4}<x<10^{-3}\,,\nonumber\\
H_{7}:\ & -10^{-7}<x<10^{-3}\,,\nonumber\\
H_{9}:\ & -10^{-9}<x<10^{-6}\,,\nonumber\\
H_{11}:\ & -10^{-11}<x<10^{-8}\,.
\end{align}

Another easy test can be done for the case when we deform all $\zeta(3)\rightarrow \zeta(3)+x$, including any $\zeta(3)^k\rightarrow (\zeta(3)+x)^k$, for all $k$. Unlike the first test, this will put bounds on $\zeta(3)$ purely in terms of other zeta values. In this case we also get an increase in precision with constraint order:
\begin{align}
H_{5}:\ &-10^{-4} < x < 10^{0}\,,\nonumber\\
H_{7}:\ &-10^{-5} < x < 10^{-1}\,,\nonumber\\
H_{9}:\ &-10^{-6} < x < 10^{-2}\,,\nonumber\\
H_{11}:\ &-10^{-7} < x < 10^{-3}\,,
\end{align}
suggesting we can indeed fix $\zeta(3)$ to arbitrary degree. 

While far from a test of our conjecture, these simple setups demonstrate the constraining power of Hankel matrices. If the full conjecture is indeed correct, regardless of the connection to the string amplitude, it implies we can in principle compute any odd zeta value with arbitrary precision purely from even zeta values.

\subsection{Combined constraints for  bicolor ordered amplitude}
\begin{figure}[h!]
  \centering
  \begin{subfigure}[b]{0.4\linewidth}
    \includegraphics[scale=0.4]{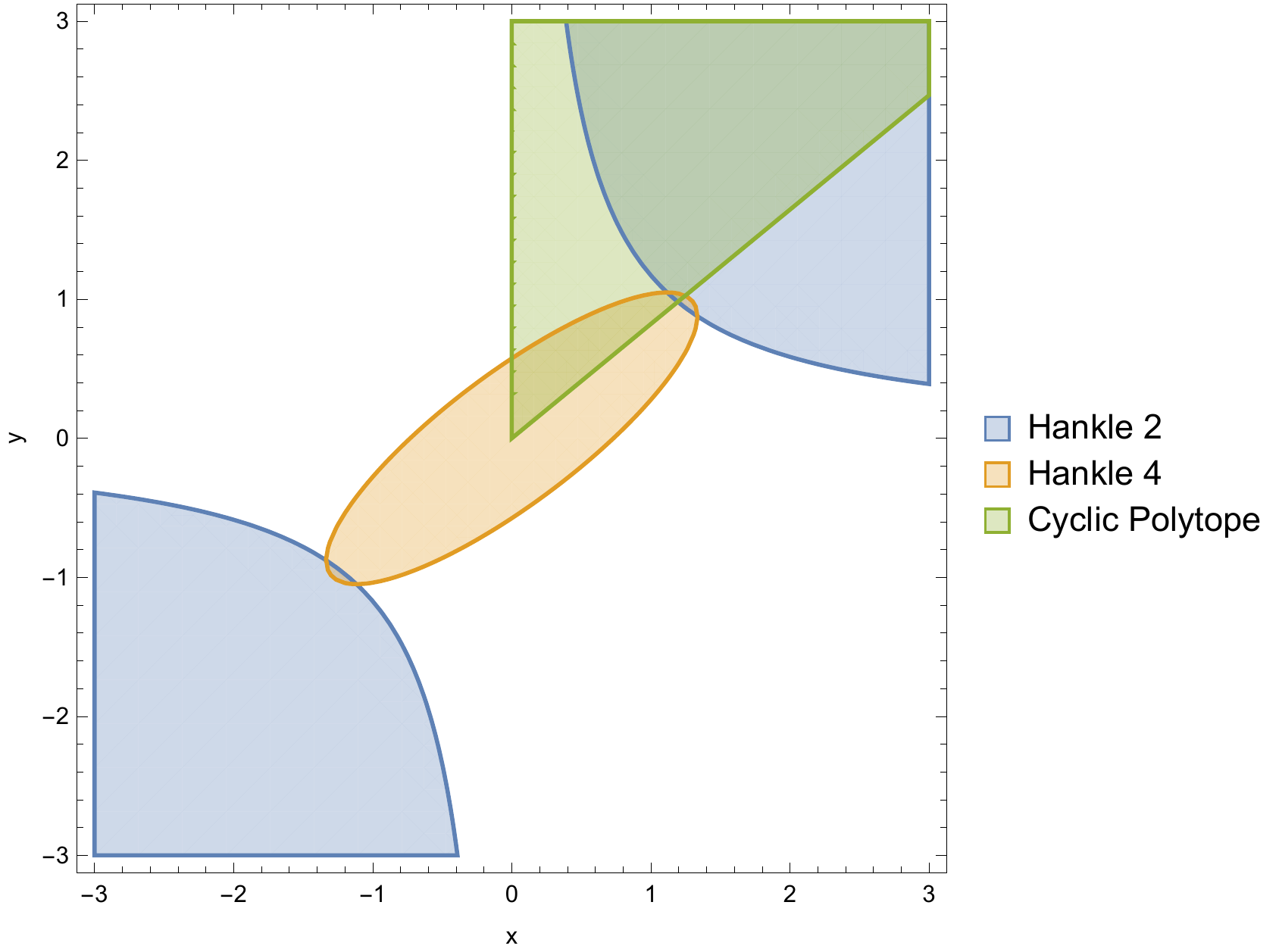}
  \caption{Cyclic Polytope+Hankel}
  \label{zha}
  \end{subfigure}
  \begin{subfigure}[b]{0.3\linewidth}
    \includegraphics[scale=0.28]{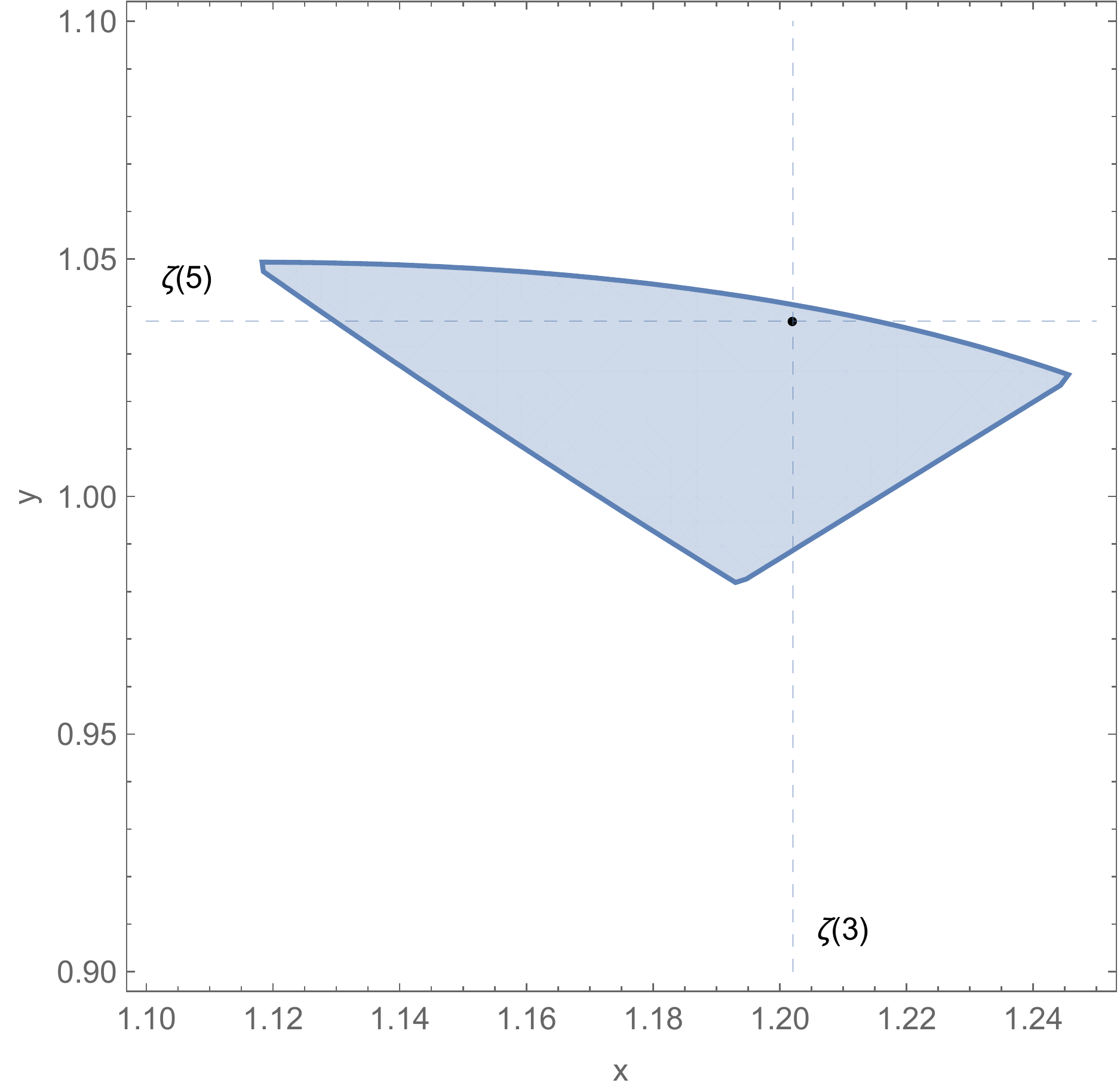}
  \caption{Intersection}
  \label{zcb}
  \end{subfigure}
  \caption{Region of parameter space for bi-color amplitude fixed by monodromy and unitarity}
\end{figure}

For the bicolor amplitude (\ref{eq:ALE}), the Laurent coefficients $g_{i,j}$ are constrained by the combination of unitarity and monodromy in a similar fashion. In this subsection we will show the result for the combined constraints on $(g_{2,0},g_{4,0})=(x,y)$ up to $k{=}5$. As shown in (\ref{eq:ALE}), the monodromy constraints read:
\begin{align}
k=0:&\quad\ g_{0,0} = 0\,,\nonumber\\
k=1:&\quad\ g_{1,0} =g_{1,1}=\zeta\left(2\right)\,,\nonumber\\
k=2:&\quad\ g_{2,1} = 2g_{2,2} =2g_{2,0}\,,\nonumber\\
k=3:&\quad\ g_{3,0} = g_{3,3} = \zeta\left(4\right)\,,\quad\ g_{3,1} = g_{3,2} = \frac{5}{4}\zeta\left(4\right)\,,\nonumber\\
k=4:&\quad\ g_{4,1} = g_{4,3} =- \zeta\left(2\right)g_{2,0}+3g_{4.0}\,,\quad\ g_{4,2} =  -2\zeta\left(2\right)g_{2,0}+4g_{4,0}\,.
\label{eq:zmono3} 
\end{align}
Thus the monodromy plane can be identified with the plane spanned by $(g_{2,0},g_{4,0})=(x,y)$. The cyclic polytope $\ref{cpc}$ expanded on the monodromy plane will give us some non-relevant conditions, which we do not list here. One nontrivial constraint is that the components of $\vec{g}_{4}$ should be positive. The positivity of the third component in $\vec{g}_4^{(3)}=g_{4,2}/g_{4,0}$ can be expanded on the monodromy plane and implies:
\eq
\frac{g_{4,2}}{g_{4,0}}>0\quad \longrightarrow\quad \frac{y}{x}>\frac{\pi^2}{12}\label{eq:ZC4}\,.
\eqe
As we know $g_{5,0}=\zeta\left(6\right)$ from monodromy, we can include positive determinant constraints for Hankel matrices ($\ref{hmc}$) containing $g_{5,0}$ 
\begin{equation*}
H_1 =
\begin{pmatrix}
g_{1,0} & g_{2,0}\\
g_{2,0} & g_{3,0}
\end{pmatrix},
H_2 =
\begin{pmatrix}
g_{2,0} & g_{3,0}\\
g_{3,0} & g_{4,0}
\end{pmatrix},
H_3 =
\begin{pmatrix}
g_{2,1} & g_{3,1}\\
g_{3,1} & g_{4,1}
\end{pmatrix}
\end{equation*}
\begin{equation}
H_4 =
\begin{pmatrix}
g_{1,0} & g_{2,0} & g_{3,0}\\
g_{2,0} & g_{3,0} & g_{4,0}\\
g_{3,0} & g_{4,0} & g_{5,0}
\end{pmatrix},
H_5 =
\begin{pmatrix}
g_{3,0} & g_{4,0}\\
g_{4,0} & g_{5,0}
\end{pmatrix},
H_6 =
\begin{pmatrix}
g_{1,0} & g_{3,0}\\
g_{3,0} & g_{5,0}
\end{pmatrix}\label{eq:ZH4}\,.
\end{equation}
The product Hankle matrices $\ref{phc}$ are also not relevant in this case. We summarize all the nontrivial constraints in (\ref{eq:zmono3}, \ref{eq:ZC4}) and the matrices $\det\left(H_i\right)>0$ in (\ref{eq:ZH4}) in Figure \ref{zha}. 
As in the single color case, the parameters are confined to a small region, shown in Figure \ref{zcb}.

\subsection{Closed string EFT from KLT}
\begin{figure}[h!]
  \centering
    \includegraphics[scale=0.4]{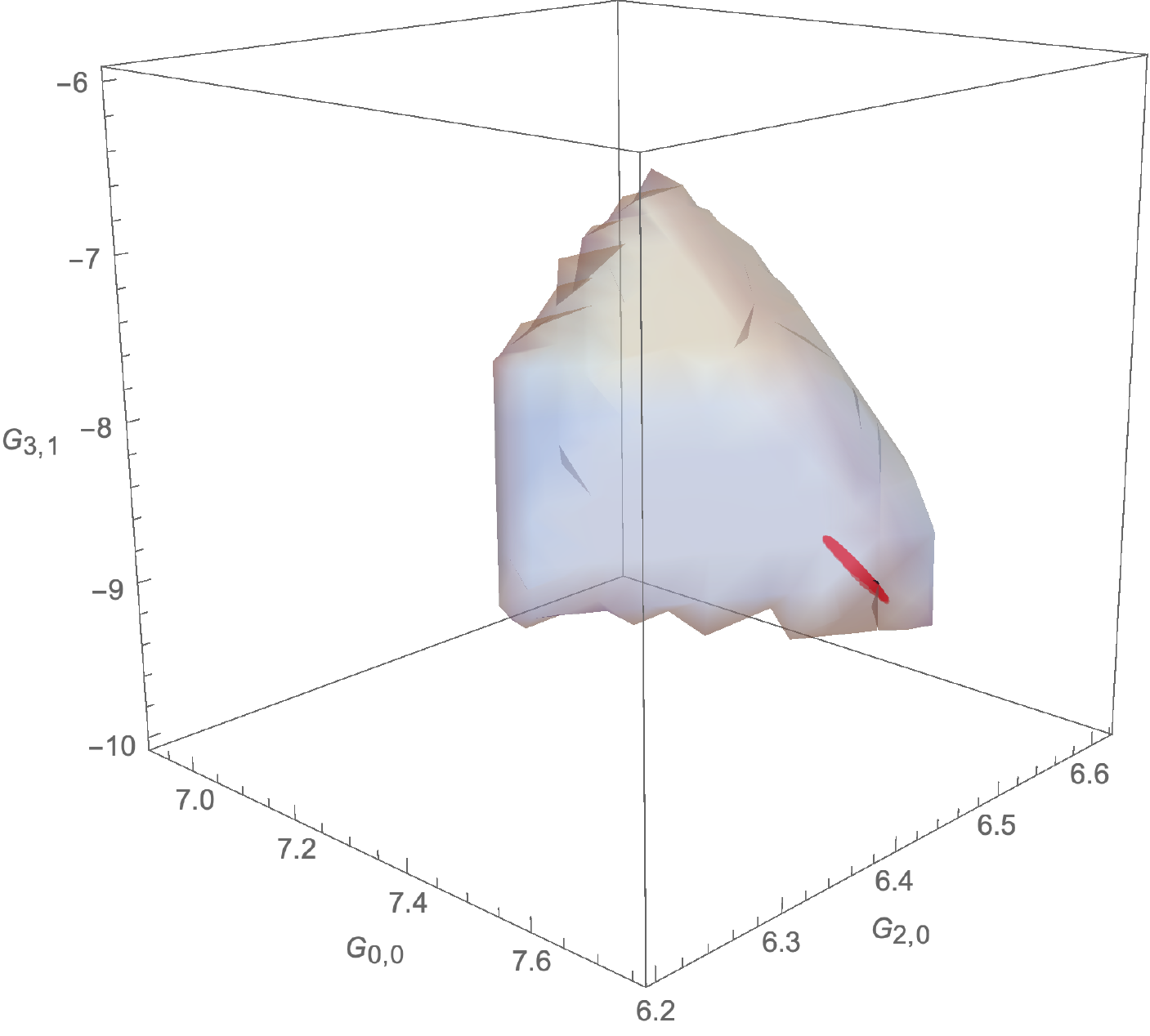}
  \caption{Closed string allowed region in d=26.}\label{fig:cs3d}
\end{figure}

Given the monodromy relation of the open string amplitude, the closed string KLT relation~\cite{Kawai:1985xq} (see also \cite{Stieberger:2009hq,BjerrumBohr:2009rd}) can be viewed as its corollary. Thus once we obtain the open string EFT, we can straightforwardly carve out the closed string EFT by projection through the KLT kernel. For example, combining (\ref{eq:ALE2}) and  (\ref{eq:agkq}) the projection through KLT yields the following:
\begin{align}
M_{\textrm{closed string}}\left(s,t\right) &=  A\left(s,t\right)\sin\left(\pi s\right)A\left(s,-s-t\right)\nonumber \\
 & =\frac{-\pi}{st\left(s+t\right)}-2\pi g_{1,0}-2\pi g_{3,0}\left(s^{2}+st+t^{2}\right)\nonumber \\
 & -\pi\left(\frac{\pi^{6}}{630}+g_{1,0}^{2}-2g_{4,1}\right)\left(s+t\right)st+2\pi g_{5,0}\left(s^{2}+st+t^{2}\right)^{2}+\ldots\label{eq:doublecopy}
\end{align}
This can be compared to the EFT expansion 
\begin{align}
M\left(s,t\right)=\frac{-\pi}{st\left(s+t\right)}+\sum_{i,j=0}^{\infty}G_{i,j}s^{i-j}t^{i}\,,
\end{align}
where  the coefficients $G_{ij}$ are linearly related to $g_{i,j}$:
\begin{align}
\nonumber G_{0,0} & =2\pi g_{1,0},\quad G_{2,0}=G_{2,1}=G_{2,2}=2\pi g_{3,0},\\
\nonumber G_{3,1} & = G_{3,2} =-\pi\left(\frac{\pi^{6}}{630}+g_{1,0}^{2}-2g_{4,1}\right),\quad G_{3,0}=G_{3,3}=0\\
G_{4,0} & =G_{4,4}=2\pi g_{5,0},\quad G_{4,1}=G_{4,4}=4\pi g_{5,0},\quad G_{4,2}=6\pi g_{5,0}\,.
\end{align}
The shape of the allowed region of independent $G_{i,j}$'s follows
from the allowed region of monodromy free parameters $g_{1,0}$, $g_{3,0}$, $g_{4,1}$, etc.
For example, the allowed region for the $G_{0,0}$, $G_{2,0}$ and $G_{3,1}$
is shown in Figure \ref{fig:cs3d}, which is a straightforward coordinate transformation
of the region in Figure \ref{fig:U43Dh1} which we used $(x,y,z)=(g_{1,0},g_{3,0},g_{4,1})$.

The EFThedron constraint for non-ordered amplitude is much more intricate and not fully explored. However, in the forward limit, the Hankel matrix positivity is a direct carbon copy of the color ordered ones. Now the double copy in the forward limit takes the form:
\begin{align}
 \nonumber & \lim_{t\rightarrow0}A\left(s,t\right)\left(\pi t\right)A\left(-s,t\right)\\
= & \left(\frac{1}{\left(\alpha'\right)^{2}st}+\sum_{i,j=0}^{\infty}g_{i,0}s^{i}\right)\left(\pi t\right)\left(\frac{1}{\left(\alpha'\right)^{2}\left(-s\right)t}+\sum_{i,j=0}^{\infty}g_{i,0}\left(-s\right)^{i}\right)\,.
\end{align}
Only the contribution from odd powers of $s$ survives and the regular terms in ${\cal M}\left(s,t\right)$ read:
\begin{align}
-2\pi\left(g_{1,0}+g_{3,0}s^{2}+g_{5,0}s^{4}+...\right)\,.
\end{align}
Thus we see that the Hankel positivity of the closed string amplitude is simply a subset of the open string ones, and the KLT kernel can be viewed as a positivity preserving projection. See also a similar discussion in \cite{Green:2019tpt}.



\section{Conclusions and outlook}
In this paper, we study the interplay of consistency conditions for a space-time S-matrix, and a CFT four-point correlation function. This can be viewed as an initial step towards an on-shell approach to carving out the string landscape. By considering a worldsheet type integral representation for the four-point amplitude, we show that consistent factorization of the S-matrix forces the integrand to be given by linear combinations of SL(2,R) conformal blocks. Unitarity of the S-matrix for positive conformal weights carves out a subregion within the space of linear combinations, where we demonstrate that the Virasoro block appears at a kink in the boundary of allowed solutions. In the cases shown, this criteria is sufficient to analytically define the Virasoro combination. Thus Virasoro symmetry emerges from the consistency of the space-time S-matrix. 

Note that a characteristic property of $\chi_q$, defined as coefficients in the linear combination of global SL(2,R) blocks for the Virasoro block, is the presence of poles associated with the null states. If instead we restrict ourselves to a polynomial ansatz,  in general we can only cover a subspace of the allowed region, and it is only when we allow for a rational ansatz can we reach the boundary. Thus the presence of null state poles is crucial in putting us on the boundary. It will be interesting to understand more deeply from the S-matrix point of view the necessity for the presence of these null states. 

In the opposite direction, we consider open string correlators with overall monodromy, which arrises for general flat-space amplitudes. We demonstrate that the resulting monodromy relations allow for three isolated solutions, each enforcing algebraic identities amongst the low energy couplings. Thus the low energy description of string theory amplitudes corresponds to the intersection of the ``monodromy plane" with the EFThedron. For the monodromy plane that arises for usual flat space amplitudes, we show that the intersection space is an isolated island, whose area rapidly converges as higher derivative order constraints are taken into account. This leads us to conjecture that the four-point open superstring amplitude is completely fixed by the geometry of the intersection between the monodromy plane and the EFThedron. We present the result for the same investigation on bicolor monodromy.

Since the intersection geometry is infinite dimensional in nature, it will be desirable to have a continuous limit description. Note that this is reminiscent of the CFT bootstrap, where the initial derivative truncation gave way to more efficient analytic functionals~\cite{Mazac:2018mdx}. The initial step would be then to have a continuous definition for the Hankel matrix bounds. A recent proposal of positive functionals is a very promising direction~\cite{Riva}.

The constrained space of open string EFT couplings naturally leads to a constrained space for closed strings couplings through KLT relations. We find that so long as the open string couplings reside in the EFThedron, the closed string image automatically satisfies all Hankel-type bounds. However, it is well known that the EFThedron for general permutation invariant theories is much more intricate than the color ordered ones \cite{Green:2019tpt}. It will be interesting in the future to see if the KLT kernel always projects the intersection geometry of the open string \textit{inside} the general EFThedron. It is clearly desirable to understand what statements can be made for the monodromy of general  string compactifications, and study how to modify our approach to cases where instead of universal monodromy, the open string amplitude is given as sum of blocks with distinct yet understood monodromy phases. Finally, it was recently shown that the string EFT expansion can be expressed in terms of just a few modified color-kinematic building blocks \cite{Carrasco:2019yyn}. It would be interesting to understand how monodromy relates the color-kinematic solutions at different mass dimension, and if monodromy-compatible solutions themselves are amenable to a direct bootstrap procedure.


\section{Acknowledgement}
We would like to thank Nima Arkani-Hamed, Shu-Heng Shao and Pierre Vanhove for discussions at the early stages of this work,  Chi-Ming Chang for enlightening discussions, Yang Zhang for helping improve efficiency of the algorithm, and John Joseph Carrasco and Congkao Wen for comments on the draft. Y-h Wang is supported by MoST Grant No. 108-2811-M-002-535. Y-t Huang and J-y Liu are supported by MoST Grant No. 106-2628-M-002-012-MY3. Y-t Huang is also supported by Golden Jade fellowship. L Rodina is supported by the European Research Council under ERC-STG-639729, Strategic Predictions for Quantum Field Theories .

\bibliography{PosCFT}
\bibliographystyle{utphys}

\end{document}